\documentclass[twocolumn,english,aps,prx,notitlepage,superscriptaddress]{revtex4-2}

\usepackage[T1]{fontenc}
\usepackage[latin9]{inputenc}
\usepackage{graphicx}
\usepackage{amsmath}
\usepackage{physics}
\usepackage{float}
\usepackage{bm}
\usepackage[colorlinks = true,
linkcolor = blue,
urlcolor  = blue,
citecolor = blue,
anchorcolor = blue]{hyperref}
\usepackage[usenames, dvipsnames]{color}

\usepackage[final]{changes}

\begin{document}

\sloppy
\title{Magnetic order, magnons, and crystal fields in van der Waals CeSiI}

\author{Wolfgang Simeth}
\email{wsimeth@lanl.gov}
\affiliation{MPA-Q, Los Alamos National Laboratory, Los Alamos, New Mexico 87545, USA}

\author{Connor A. Occhialini}
\affiliation{Department of Physics, Columbia University, New York, New York 10027, USA}
\affiliation{National Synchrotron Light Source II, Brookhaven National Laboratory, Upton, NY 11973, USA}

\author{Michael E. Ziebel}
\affiliation{Department of Physics, Columbia University, New York, New York 10027, USA}
\affiliation{Department of Chemistry, Columbia University, New York, NY, USA}

\author{Nethmi W. Hewage}
\affiliation{Department of Chemistry, Columbia University, New York, NY, USA}

\author{Sabrina J. Li}
\affiliation{Computational Physics Division, Los Alamos National Laboratory, Los Alamos, New Mexico 87545, USA}
\affiliation{Center for Nonlinear Studies, Los Alamos National Laboratory, Los Alamos, New Mexico 87545, USA}

\author{Daniel Pajerowski}
\affiliation{Neutron Scattering Division, Oak Ridge National Laboratory, Oak Ridge, TN 37831, USA}

\author{Taehun Kim}
\affiliation{Department of Physics, Columbia University, New York, New York 10027, USA}
\affiliation{National Synchrotron Light Source II, Brookhaven National Laboratory, Upton, NY 11973, USA}

\author{Ben Zager}
\affiliation{National Synchrotron Light Source II, Brookhaven National Laboratory, Upton, NY 11973, USA}
\affiliation{Department of Condensed Matter Physics and Materials Science, Brookhaven National Laboratory, Upton, New York 11973, USA}

\author{Jonathan Pelliciari}
\affiliation{National Synchrotron Light Source II, Brookhaven National Laboratory, Upton, NY 11973, USA}

\author{Kipton Barros}
\affiliation{Theoretical Division and CNLS, Los Alamos National Laboratory, Los Alamos, New Mexico 87545, USA}

\author{Daniel Rehn}
\affiliation{Computational Physics Division, Los Alamos National Laboratory, Los Alamos, New Mexico 87545, USA}

\author{Abhay N. Pasupathy}
\affiliation{Department of Physics, Columbia University, New York, New York 10027, USA}
\affiliation{Department of Condensed Matter Physics and Materials Science, Brookhaven National Laboratory, Upton, New York 11973, USA}

\author{Valentina Bisogni}
\affiliation{National Synchrotron Light Source II, Brookhaven National Laboratory, Upton, NY 11973, USA}

\author{Xavier Roy}
\affiliation{Department of Chemistry, Columbia University, New York, NY, USA}

\author{Allen Scheie}
\affiliation{MPA-Q, Los Alamos National Laboratory, Los Alamos, New Mexico 87545, USA}

\date{\today}

\begin{abstract}
We report neutron, X-ray absorption, and resonant X-ray spectroscopy of magnetic excitations in the new heavy-fermion van-der-Waals superconductor CeSiI. 
We determined effective Hamiltonians and ground states of crystal electric fields and magnons.
Isotropic Heisenberg interactions on a quasi two dimensional lattice, including ferromagnetic nearest-neighbor exchange as the dominant interaction, provide an excellent account to the low-energy measured dynamics and stabilize a co-rotating spin cycloid. 
Our study provides the basis to model CeSiI from first principles, thereby laying the ground for microscopic understanding of heavy-fermion physics, their unconventional superconductivity, and quantum criticality.
\end{abstract}

\maketitle

Two-dimensional materials are famous for enhancing electronic correlations, thereby creating an exceptionally fertile platform for quantum phenomena~\cite{2005_Kane_PhysRevLettb,2018_Cao_Natureb,2016_Lee_NanoLett,2016_Mak_NaturePhoton,2021_Vano_Nature,2011_Mizukami_NaturePhys}. Van-der-Waals (vdW) materials are particularly interesting in this regard combining the advantages of two-dimensional physics~\cite{KOSTERLITZ1978371} with those of bulk materials~\cite{2017_Gong_Nature,2017_Huang_Nature,2018_Burch_Nature,2024_Ziebel_NanoLett}. 
The ability to exfoliate these states to monolayers~\cite{2017_Gong_Nature,2017_Huang_Nature,2025_Thomas_} and to reassemble them to heterostructures~\cite{2016_Novoselov_Science,2016_Liu_NatRevMater} opens up almost endless possibilities for advanced functionalities~\cite{2017_Jariwala_ACSPhotonics,2025_Zotev_NatPhoton}.
In addition, vdW materials with reduced dimensionality present a unique opportunity to study quantum phases in a theoretically tractable regime.

In this context, the two-dimensional van der Waals (vdW) heavy fermion magnet CeSiI represents an important breakthrough~\cite{Jang2022,posey2024CeSiI,2026_Shi_}. CeSiI is the first vdW heavy fermion superconductor, features strange-metal behavior and multiple signs of two-dimensional physics, and seems further to fulfill all the requirements for $p$-wave superconducting pairing~\cite{2026_Shi_}, a promising route towards topological superconductivity~\cite{2016_Kallin_RepProgPhys,2015_Sarma_npjQuantumInf,2017_Karzig_PhysRevB,2017_Sato_RepProgPhysa}. Because it can be exfoliated down to few layers, its quantum state can be probed with high precision (and potentially be incorporated into heterostructures to tune its properties~\cite{2016_Liu_NatRevMater,2016_Novoselov_Science}). Perhaps most important, quantum calculations are more tractable in two than in three dimensions~\cite{2023_Yadav_Matter,Stoudenmire2012_DMRG,pan2025quantum}. This makes CeSiI a key target not only for potential applications but also as a new pathway to answer long-standing questions about heavy fermions, such as the microscopic mechanisms behind unconventional superconductivity and quantum criticality~\cite{2007_Monthoux_Naturea}.

\begin{figure}
    \includegraphics[]{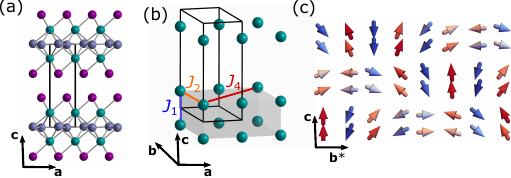} 
    \caption{(a) Trigonal crystal structure of CeSiI and conventional hexagonal unit cell (black skeleton, basis vectors $\bm{a}$, $\bm{b}$, and $\bm{c}$). Shown are two pairs of Ce (green spheres) and the surrounding Si (gray) and I layers (purple). (b) Quasi two dimensional spin Hamiltonian that accounts for the spin dynamics in CeSiI. Shown are 1st, 2nd, and 4th nearest neighbor bonds as introduced in the Hamiltonian (Eq.~\ref{Equation1}). (c) Magnetic ground state in agreement with our data and model. Shown are magnetic moments carried by Ce in the plane perpendicular to \replaced{$\bm{b}$}{$\bm{a}$}.}
    \label{fig:Figure1}
\end{figure}

CeSiI reveals strong correlations with a large Sommerfeld coefficient $\gamma=125$\,mJ/mol$\cdot$K$^2$ and large effective mass from quantum oscillations~\cite{posey2024CeSiI}. Its two-dimensional vdW intermetallic structure is made of almost perfectly flat honeycomb layers of Si, sandwiched between triangular planes of Ce [see Fig.~\ref{fig:Figure1}(a)]. Each intermetallic slab is capped with iodine. Photoemission spectroscopy revealed a two-dimensional electronic structure~\cite{posey2024CeSiI}, and  
hybridization between conduction and $f$-electrons is highly anisotropic, likely as a consequence of dimensional reduction~\cite{2025_Turkel_NatPhysa}. Long-ranged antiferromagnetic order of Ce ions emerges below $T_{\mathrm{N}}=7.5\,$K with incommensurate ordering wave-vector $\bm{k}=(0.28,0,0.19)\,$r.l.u. (reciprocal lattice units); however structure refinements were unable to distinguish between three candidate ground states: co-rotating cycloid, counter-rotating cycloid, or a spin density wave~\cite{2021_Okuma_PhysRevMater}.

\begin{figure*}
    \includegraphics[width=0.99\textwidth]{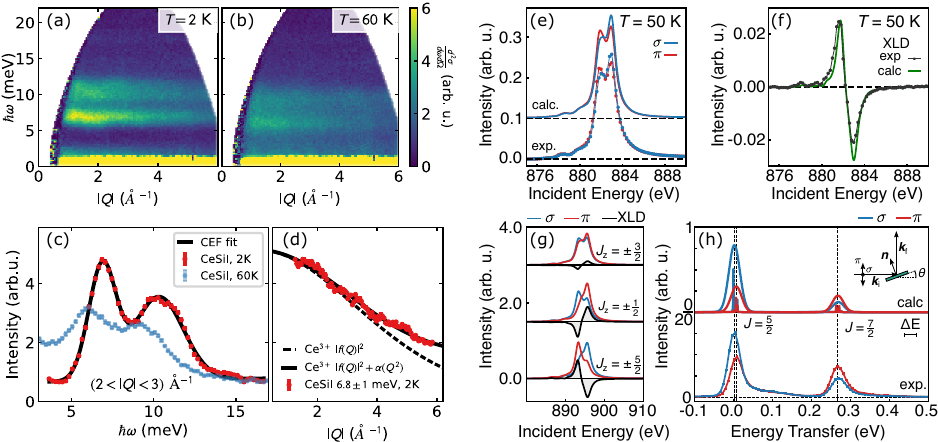} 
    \caption{Crystal electric field (CEF) analysis of CeSiI, inelastic neutron scattering, X-ray absorption spectroscopy, and resonant inelastic X-ray scattering. Panels (a) and (b) show the powder neutron spectrum with $E_i = 25$~meV at 2~K and 60~K. Panel (c) shows the fitted model compared with the two measured CEF levels. At 60~K the crystal field levels shift to lower energies. Panel (d) shows the $Q$-dependence of the lowest mode intensity, which matches the Ce$^{3+}$ form factor plus a small $Q^2$ phonon background.\cite{Squires}   
    (e) Ce $M_5$-edge XAS ($881.9\,$eV) recorded in TEY with incident $\sigma$ (blue) and $\pi$ (red) polarizations at incident angle $\theta = 20\,$deg and $T = 50$ K. (f) XLD spectrum determined from data in (e) by subtracting $\pi$ from $\sigma$. Solid green, blue and green lines in (e,f) are fits to the data (see text), circles denote recorded data. (g) Calculated XLD for each $\ket{J_z}$ term within the $J = 5/2$ manifold. (h) Ce $M_5$-edge RIXS spectrum ($E_i=881.9\,$eV) recorded with $\Delta E = 31$ meV resolution at $T=50\,$K, incident angle $\theta = 22.5\,$deg and scattering angle $2\theta = 90\,$deg with incident $\sigma$ (blue) and $\pi$ (red) polarization. The schematics illustrates the scattering geometry. The data are compared to calculations from the same model as fit to the XLD in (e,f) (solid lines) with and without convolution by the Gaussian resolution function (broad and sharp spectra, respectively). Dashed vertical lines indicate maxima in the calculated spectra. The dashed horizontal line indicates zero intensity.
    }  
    \label{fig:Figure2}
\end{figure*}

CeSiI is therefore a unique candidate material to explore strongly-correlated phenomena from first principles and key both for long-standing fundamental questions and in the context of applications.
To resolve and effectively model the unusual emergent properties of CeSiI however, it is crucial to understand its single-ion crystal electric field (CEF) state, magnetic ground state, and dominant magnetic interactions. 
In this letter, we combine neutron spectroscopy, X-ray absorption spectroscopy (XAS), and resonant inelastic X-ray scattering (RIXS) to address all three questions. We unambiguously determine the CEF ground state, determine a minimal magnon model that captures the magnetic order and dynamics in CeSiI, and identify a co-rotating spin cycloid [Fig.~\ref{fig:Figure1}(c)] as the magnetic ground state.


We measured the powder inelastic neutron spectrum of 1.9~g CeSiI using the CNCS spectrometer \cite{CNCS} at Oak Ridge National Laboratory's Spallation Neutron Source \cite{mason2006spallation}. 
We further performed X-ray absorption spectroscopy (XAS) and resonant inelastic X-ray scattering (RIXS) on single-crystal CeSiI at the Ce $M_5$-edge ($3d \to 4f$) at the 2-ID (SIX) beamline at the National Synchrotron Light Source II, Brookhaven National Laboratory~\cite{2016_Dvorak_RevSciInstrum}. XAS measurements were recorded in total electron yield (TEY), with incident linear X-ray polarization parallel (denoted $\pi$) and perpendicular ($\sigma$) to the plane defined by incident X-rays, $\bm{k}_i$, and sample normal, $\bm{n}$. For X-ray measurements, samples were cleaved in-situ in ultrahigh vacuum ($P<10^{-8}\,$mbar) at room temperature. TEY XAS is shown in figure~\ref{fig:Figure2}(e). Ultrahigh resolution RIXS spectra ($\Delta E=31\,$meV, $E_i=881.9\,$eV) are presented in Fig.~\ref{fig:Figure2}(h), lower panel.
Additional details of experiments are given in the Supplementary Information (SI)~\cite{SuppMat}. 


At high energies in the neutron data, there are two visible modes at 6.88(3)\,meV and 10.31(5)\,meV at $T=2$~K [Fig.~\ref{fig:Figure2}(a-c)]. Their intensity follows the magnetic form factor Ce$^{3+}$ $Q$-dependence [Fig. \ref{fig:Figure2}(d)], showing they are magnetic in origin. A weak dispersion is present in the low-energy mode at 2~K (6.88\,meV). However, the fact that these two modes persist well above $T_N$ indicates these are CEF excitations. The width of the two modes along the energy axis is considerably larger than instrumental resolution, $\Delta E= 1.80\,$meV; this can be explained by the vibronic coupling between CEF states and phonons, as calculated using density functional theory (see SI \cite{SuppMat}).

We model and fit these excitations using the \texttt{PyCrystalField} \cite{PyCrystalField} software and start with the CEF Hamiltonian written as:
\begin{equation}
\mathcal{H}_{CEF} = \sum_{n,m} B_n^m O_n^m  \, ,
\label{Equation1}
\end{equation}
where $B_n^m$ are scalar CEF parameters and $O_n^m$ Stevens operators \cite{Stevens1952}. Because of the three-fold rotation symmetry at the Ce site ($C_{3\nu}$), only three parameters are nonzero for the $J=5/2$ multiplet: $B_2^0$, $B_4^0$, and $B_4^3$. Fitting to the experimental data yields three possible Hamiltonians and ground state wave-functions (see SI \cite{SuppMat}) that describe the neutron data equally well but which display different sequences of excited state levels.

To adjudicate between these three possibilities, we turn to the XAS (circles) in Fig. \ref{fig:Figure2}(e). The pronounced double-peak centered around 882\,eV corresponds to the Ce $M_5$-edge. The difference between the $\sigma$ (blue) and $\pi$ (red) channels constitutes the X-ray linear dichroism (XLD) and is presented in panel (f).
The magnitude and sign of XLD reflect the relative contributions of the three Kramer's doublets ($J_z =  \pm 5/2$, $\pm 3/2$, $\pm 1/2$) to the overall Ce ground state wavefunction~\cite{2008_Hansmann_PhysRevLett,2015_Willers_ProcNatlAcadSci,2024_Christovam_PhysRevLett}. To quantify this, we performed Ce M5 XAS and RIXS calculations using the Quanty software~\cite{2012_Haverkort_PhysRevB}.
As a reference, panel (g) shows the calculated spectra for the two incident polarizations and the three bare states $J_z =  \pm 3/2$, $\pm 1/2$, and $\pm 5/2$.
The sign of the XLD immediately indicates that $\pm 5/2$ is the dominant ground state contribution. Meanwhile, the $C_{3n}$ symmetry of Ce allows for finite mixing of $\pm 5/2$ and $\pm 1/2$.
We inferred the precise CEF induced mixing by fitting a superposition of these two states to XLD data, while accounting for finite temperature effects using the mode energies determined by the INS spectrum. The fit yields a ground state wavefunction of $\psi_0{\pm} = -0.66(2) \ket{\mp \frac{1}{2} }+ 0.77(2) \ket{\pm \frac{5}{2} } $. Calculations (calc) compared to experimental (exp) data are shown in panel (f) (gray datapoints vs green curve). The two polarization channels also show close quantitative agreement with calculations [panel (e) calc vs exp], also confirmed on a separate sample reported in the SI \cite{SuppMat}. The mixing of $ \ket{\mp \frac{5}{2} }$ and $ \ket{\mp \frac{1}{2} }$ in the ground state wave-function reflects significant quantum tunneling.

The fitted ground state result agrees to within uncertainty with one of the three solutions from neutron data: $\psi_0{\pm} = -0.66(5)\ket{\mp \frac{1}{2} }+ 0.75(5)\ket{\pm \frac{5}{2} } $, from the fitted CEF parameters $B_2^0 = -0.09(8)$~meV, $B_4^0 = -0.0055(9)$~meV, $B_4^3 = -0.546(11)$~meV.  
The $g$-tensor, characterized by $g_{xx} = g_{yy} = 1.1(2)$ and $g_{zz} = 2.0(4)$, reflects only weak anisotropy of the effective spin. 

Resonant inelastic x-ray scattering (RIXS) on the Ce $M_5$-edge further supports the ground state wave function identified with neutrons and XAS. 
The RIXS spectra in the lower panel (g) exhibit two peaks around 0-10 meV and 263 meV, representing $f\to f$ transitions to the spin--orbit-split $J = \frac{5}{2}$ and $J = \frac{7}{2}$ manifolds~\cite{2016_Amorese_PhysRevB}, respectively. Due to polarization selection rules, strong dichroism is observed between $\sigma$ and $\pi$ polarization, with the spin--orbit coupling (SOC)-split peaks exhibiting a $\sim6$~meV difference in their energy splitting. Calculations of the RIXS intensity within the same model fit to the XLD and neutron data give a good agreement [Fig. \ref{fig:Figure2}(g, upper panel)], while also confirming the absence of higher energy excitations related to the CEF. Meanwhile, the experimentally determined SOC ($\zeta\sim75.69\,$meV) compared to the CEF energy scale ($\sim10\,$meV) yields a nearly pure $J = \frac{5}{2}$ ground state within our single-ion multiplet model. 

\begin{figure*}
    \includegraphics[width=
    \textwidth]{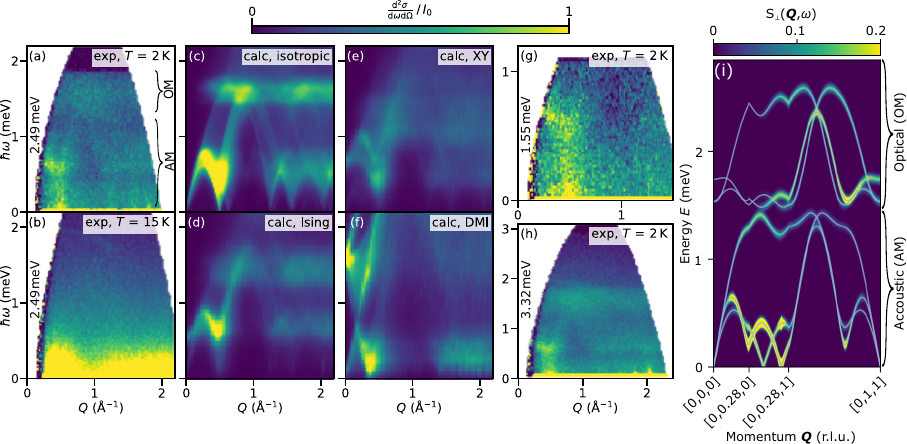} 
    \caption{Magnon dynamics in CeSiI. (a,b) Powder spectrocopy data of CeSiI taken at $T=2\,$K and $T=15\,$K, respectively. The incident neutron energy used is indicated with the vertical label on the left. In (a) we see the projected scattering of acoustic magnon branches (AB) including a pronounced Goldstone mode as well as optical branches (OB). (c) Powder spectrum calculated (calc) with the isotropic Heisenberg model in Eq.~\ref{Equation2} (with exchange $J_1$--$J_{12}$). (d,e) Calculated spectra with anisotropic nearest neighbor exchange $J_1$ and isotropic exchange $J_2$--$J_4$ (see SI for further information). We considered an anisotropic exchange matrix $\mathrm{J}$ where (d) the XY-component and (e) the Z components are reduced to $\frac{1}{2}\cdot J_1$. (f) Simulations for a model with $J_1$ and $J_2$ as well as DM interactions of strength $1.19\,$meV along $\left\langle100\right\rangle$ bonds. Further details are provided in the SI. $I_0$ in panel (f) is half of the value chosen for panels (c)-(e). (g,h) Further powder spectroscopy data taken at $T=2\,$K but with other incident energies. (i) Three dimensional magnon band structure calculated from Eq.~\ref{Equation2}. The value $I_0$ is equal for panels (a) and (b).
    }  
    \label{fig:Figure3}
\end{figure*}

Having identified the single-ion magnetic wavefunction, we now turn to the collective excitations below 3~meV in Fig.~\ref{fig:Figure3}. 

The main features of the experimental (exp) powder inelastic neutron scattering spectrum in Fig.~\ref{fig:Figure3}(a) are:
First, a pronounced (gapless) Goldstone mode (GM) is seen at $|\bm{Q}_0|=0.49$\AA$^{-1}$, which corresponds to the magnetic zone center $\bm{Q}_0=(0.28,0,0.19)$, as well as related momentum transfers of higher Brillouin zones. 
Second, intensity of acoustic magnon (AM) branches, which include the GM, are seen at energies below 1.4\,meV and most pronounced on an almost flat line around 0.6\,meV [indicated by a curly bracket]. 
Third, optical magnons (OM) at energies between 1.4\,meV and 2.5\,meV, while most pronounced around 1.6\,meV [curly bracket in (a)]. We associate these features with magnon modes because they vanish with increasing temperature [see Fig.~3(b) in comparison to (a)]. 
In the paramagnetic state [Fig.~\ref{fig:Figure3}(b)], we further observe temperature dependent quasielastic scattering ($E=0\,$meV). At $T=15\,$K [panel (b)] the linewidth is of the order 0.5\,meV and increases monotonically in $T$~(see SI~\cite{SuppMat}). Such scattering is frequently observed in heavy-fermion materials~\cite{1981_Horn_PhysRevB,1987_Erkelens_,1988_Grier_JPhysCSolidStatePhys}. Interpreted in the context of a single-impurity model it suggests a Kondo temperature of the order $T_{\mathrm{K}}=4.6\,$K, which is significantly larger than the Fano-linewidth observed in scanning tunneling spectroscopy~\cite{posey2024CeSiI}.



We can successfully model the magnon spin dynamics with a quasi two-dimensional lattice with fully isotropic interactions, where two nearest-neighbor cerium layers (within a single vdW layer) are strongly coupled to form a co-rotating spiral ground state (see Fig.~\ref{fig:Figure1}(c)). 
The effective Hamiltonian is: 
\begin{align}
    H = \sum_{n,\left\langle i,j\right\rangle\in \mathcal{B}_n} J_{n}  \bm{S}_{i} \cdot \bm{S}_{j}  ,
    \label{Equation2}
\end{align}
where exchange $J_n$ corresponds to a set of $n$-th shortest bonds $\mathcal{B}_n$ (see SI). Three isotropic Heisenberg exchange couplings (see Fig.~\ref{fig:Figure1}(b)) are sufficient to capture the magnon dynamics seen with neutrons. These include ferromagnetic nearest neighbor exchange ($J_1=-0.93$~meV) as well as competing ferro- and antiferromagnetic, respectively, nearest and next-nearest neighbor intralayer exchange ($J_2=-0.39$~meV, $J_4=0.62$~meV), which stabilize the in-plane incommensurate wavevector $k_x=0.28$. 
The gapless nature of excitations rules out significant easy-axis single-ion anisotropy.
The scale of $J_1$ is set by the highest energy 1.6~meV mode (optical magnons), and the scale of $J_2$ is set by comparing data and theory via a linecut through $S(Q,\omega)$ at $Q=0.35\,\mathrm{\AA}^{-1}$ (see SI). The ratio $J_4/J_2$ is constrained by Luttinger-Tisza (LT) analysis~\cite{1974_Litvin_Physica}, which requires $J_4/J_2=-1.6$ to stabilize incommensurate modulation vector $k_x=0.28$ (also derived in Ref.~\cite{2024_Vijayvargia_}).

With this minimal model, which we call quasi-two-dimensional as it excludes interactions spanning the vdW-gap, we are able to capture all the main magnon features explained above. Fig.~\ref{fig:Figure3}(c) shows the results from our simulations (calc) based on linear spin wave theory (LSWT) and carried out with \texttt{Sunny} software~\cite{2025_Dahlbom_}, indicating excellent agreement between theory and experiment. 
Meanwhile, comparatively weak interactions through the vdW gap (such as competing Heisenberg exchange or DM interactions) are required to stabilize the out-of-plane magnetic ordering vector $k_z=0.19$, but that otherwise leave the powder dynamic spectrum unaffected [see SI]. Although our experiments cannot determine the precise composition of these interplanar interactions in CeSiI, we determined using global minimization on the Luttinger-Tisza method a set of candidate bonds given by $J_5$, $J_9$, $J_{11}$, and $J_{12}$ (with bond lengths between $7.7\,$\AA~and $9.5\,$\AA)  (details are found in the SI). 

The ground state of our minimal model corresponds to a co-rotating spiral. This is necessarily the case when $J_1$ is ferromagnetic and the dominant exchange (see SI \cite{SuppMat}). Although our model [Eq.~\ref{Equation2}] does not define the plane of spin rotation, neutron powder diffraction showed unambiguously that in case of a spiral with wave-vector $\bm{k}$ magnetic moments lie in the plane perpendicular to $\mathbf{a}$ (see SI)~\cite{2021_Okuma_PhysRevMater}. 

We performed a systematic search over alternative Hamiltonians that reproduce the same propagation vector but instead favor counter-rotating spin spirals. 
This may be achieved on the one hand when the nearest neighbor exchange, $\mathrm{J}_1$, is anisotropic, however Ising-like or XY-character does not reproduce the intense Goldstone mode below 0.4\,meV as seen at $\bm{Q}_0$ and at higher momentum transfers, but results in a gapped dispersion not supported by our data [see Fig.~\ref{fig:Figure3}(d) and (e) and SI]. On the other hand by local DM interactions, facilitated by Ce ions decorating locally noncentrosymmetric sites, however these also can be ruled out because the calculated magnons do not match the experimental spectrum [see Fig.~\ref{fig:Figure3}(f)].
Spin-density wave (SDW) ground states, in turn, may emerge, when quantum fluctuations are enhanced in conjunction with strong anisotropy~\cite{1977_Allen_PhysRevB,2013_Chen_PhysRevB,2022_Facheris_PhysRevLett,2025_Park_NatCommun,2010_Starykh_PhysRevB}. Although we cannot rule a SDW out, our data are excellently described by semiclassical simulations with isotropic interactions, supporting a co-rotating spiral state. 
To finally identify the modes AM and OM as acoustic and optical magnons, we compute the theoretical single-crystal magnon band structure [Fig.~\ref{fig:Figure3}(g)]. The spectrum comprises of three acoustic and three optical modes that can be clearly attributed to the intensities AM and OM, respectively. 
Complementary powder spectroscopy data, taken also at $2\,$K but with other incident energies [Fig.~\ref{fig:Figure3}(g) and (h)], confirm the gapless nature of AM modes. 

The experimental $J_1$ is ferromagnetic ($J_1<0$), opposite to the sign predicted by density functional theory~\cite{posey2024CeSiI}. However, experimentally there is no ambiguity: antiferromagnetic $J_1$ produces neither an intense Goldstone mode nor intensity of optical magnon modes seen in our data (see SI). 

We may ask, given the heavy-fermion nature of CeSiI, whether collective modes, such as Kondo resonances, are visible in the spectrum. Due to the approximate nature of our model and the powder averaging of our data we cannot say for certain---however we do note that optical magnons and the elastic line are surrounded by a relatively diffuse scattering that is broader than our model predicts [compare Figs.~\ref{fig:Figure3}(a) and (c)]. Furthermore, the optical modes around 1.7~meV are also flatter than the simulations predict. This could be associated with a resonance of the Kondo lattice and warrants further theoretical modeling.

Given the success of a minimal model to describe the magnon dynamics, we may ask whether it also reproduces the field-dependent phase diagram with steps in magnetization~\cite{posey2024CeSiI}. Classical finite-temperature magnetization calculations (shown in the SI \cite{SuppMat})~\cite{2025_Dahlbom_,2022_Dahlbom_PhysRevB} display linear field dependence, unlike experiments~\cite{posey2024CeSiI} (and unlike calculations done on a $S=1$ CeSiI model~\cite{2024_Vijayvargia_}, which our CEF fits rule out as unphysical).  
Although anisotropy and DM interactions can, in principle, produce such steps, these need to become comparatively strong, which is not supported by our data (see SI).
We note however, that genuine quantum effects, which our magnetization calculations neglect, can produce magnetization plateaus~\cite{Chubukov_1991,PhysRevB.94.075136} and which we identify as the most likely origin for the step-like magnetization.

In summary, we determined the CeSiI crystal-field Hamiltonian, the leading magnetic exchange terms, their energy scales, the magnetic ground state, and we highlighted the relevance of quantum fluctuations. 
This information (especially, the single-ion and collective magnetic ground state) is key to building effective models of CeSiI required to unravel its correlated phenomena.
Because of the computationally advantageous low-dimensionality and atomic setting of one unpaired $f$-electron, CeSiI is a promising starting point for a quantitative understanding of vdW heavy-fermion materials in terms of \textit{ab-initio} theories. We lay the ground for these studies, which present an exciting pathway for finally understanding the long-standing mystery of heavy fermions and unconventional superconductivity.

\textbf{Acknowledgements:} We acknowledge fruitful discussions with F. Ronning, J. Thompson, A. Millis, and C. S. Ong. Research on two-dimensional heavy fermion materials at Columbia University was supported by the US Department of Energy (DOE), Office of Science, Basic Energy Science, under award DE-SC0023406 (M.E.Z., N.W.H., A.N.P., X.R.). A portion of this research used resources at the Spallation Neutron Source, a DOE Office of Science User Facility operated by the Oak Ridge National Laboratory. The beam time was allocated to CNCS on proposal number IPTS-33399.1. 
AS, WS, and KB acknowledge support from the Laboratory Directed Research and Development program of Los Alamos National Laboratory under project number 20240083DR to fund travel for the neutron beamtime. Analysis and modeling of neutron data was supported by U.S. Department of Energy, Office of Basic Energy Sciences, Division of Materials Science and Engineering under project ``Quantum Fluctuations in Narrow-Band Systems.'' The XAS and RIXS investigation was supported as part of Programmable Quantum Materials, an Energy Frontier Research Center funded by the US Department of Energy (DOE), Office of Science, Basic Energy Sciences (BES), under award DE-SC0019443 (C.A.O.). This research used beamline 2-ID (SIX) of the National Synchrotron Light Source II, a U.S. Department of Energy (DOE) Office of Science User Facility operated for the DOE Office of Science by Brookhaven National Laboratory under Contract No. DE-SC0012704.


\begin{thebibliography}{55}%
\makeatletter
\providecommand \@ifxundefined [1]{%
 \@ifx{#1\undefined}
}%
\providecommand \@ifnum [1]{%
 \ifnum #1\expandafter \@firstoftwo
 \else \expandafter \@secondoftwo
 \fi
}%
\providecommand \@ifx [1]{%
 \ifx #1\expandafter \@firstoftwo
 \else \expandafter \@secondoftwo
 \fi
}%
\providecommand \natexlab [1]{#1}%
\providecommand \enquote  [1]{``#1''}%
\providecommand \bibnamefont  [1]{#1}%
\providecommand \bibfnamefont [1]{#1}%
\providecommand \citenamefont [1]{#1}%
\providecommand \href@noop [0]{\@secondoftwo}%
\providecommand \href [0]{\begingroup \@sanitize@url \@href}%
\providecommand \@href[1]{\@@startlink{#1}\@@href}%
\providecommand \@@href[1]{\endgroup#1\@@endlink}%
\providecommand \@sanitize@url [0]{\catcode `\\12\catcode `\$12\catcode `\&12\catcode `\#12\catcode `\^12\catcode `\_12\catcode `\%12\relax}%
\providecommand \@@startlink[1]{}%
\providecommand \@@endlink[0]{}%
\providecommand \url  [0]{\begingroup\@sanitize@url \@url }%
\providecommand \@url [1]{\endgroup\@href {#1}{\urlprefix }}%
\providecommand \urlprefix  [0]{URL }%
\providecommand \Eprint [0]{\href }%
\providecommand \doibase [0]{https://doi.org/}%
\providecommand \selectlanguage [0]{\@gobble}%
\providecommand \bibinfo  [0]{\@secondoftwo}%
\providecommand \bibfield  [0]{\@secondoftwo}%
\providecommand \translation [1]{[#1]}%
\providecommand \BibitemOpen [0]{}%
\providecommand \bibitemStop [0]{}%
\providecommand \bibitemNoStop [0]{.\EOS\space}%
\providecommand \EOS [0]{\spacefactor3000\relax}%
\providecommand \BibitemShut  [1]{\csname bibitem#1\endcsname}%
\let\auto@bib@innerbib\@empty
\bibitem [{\citenamefont {Kane}\ and\ \citenamefont {Mele}(2005)}]{2005_Kane_PhysRevLettb}%
  \BibitemOpen
  \bibfield  {author} {\bibinfo {author} {\bibfnamefont {C.~L.}\ \bibnamefont {Kane}}\ and\ \bibinfo {author} {\bibfnamefont {E.~J.}\ \bibnamefont {Mele}},\ }\bibfield  {title} {\bibinfo {title} {Quantum {{Spin Hall Effect}} in {{Graphene}}},\ }\href {https://doi.org/10.1103/PhysRevLett.95.226801} {\bibfield  {journal} {\bibinfo  {journal} {Physical Review Letters}\ }\textbf {\bibinfo {volume} {95}},\ \bibinfo {pages} {226801} (\bibinfo {year} {2005})}\BibitemShut {NoStop}%
\bibitem [{\citenamefont {Cao}\ \emph {et~al.}(2018)\citenamefont {Cao}, \citenamefont {Fatemi}, \citenamefont {Fang}, \citenamefont {Watanabe}, \citenamefont {Taniguchi}, \citenamefont {Kaxiras},\ and\ \citenamefont {{Jarillo-Herrero}}}]{2018_Cao_Natureb}%
  \BibitemOpen
  \bibfield  {author} {\bibinfo {author} {\bibfnamefont {Y.}~\bibnamefont {Cao}}, \bibinfo {author} {\bibfnamefont {V.}~\bibnamefont {Fatemi}}, \bibinfo {author} {\bibfnamefont {S.}~\bibnamefont {Fang}}, \bibinfo {author} {\bibfnamefont {K.}~\bibnamefont {Watanabe}}, \bibinfo {author} {\bibfnamefont {T.}~\bibnamefont {Taniguchi}}, \bibinfo {author} {\bibfnamefont {E.}~\bibnamefont {Kaxiras}},\ and\ \bibinfo {author} {\bibfnamefont {P.}~\bibnamefont {{Jarillo-Herrero}}},\ }\bibfield  {title} {\bibinfo {title} {Unconventional superconductivity in magic-angle graphene superlattices},\ }\href {https://doi.org/10.1038/nature26160} {\bibfield  {journal} {\bibinfo  {journal} {Nature}\ }\textbf {\bibinfo {volume} {556}},\ \bibinfo {pages} {43} (\bibinfo {year} {2018})}\BibitemShut {NoStop}%
\bibitem [{\citenamefont {Lee}\ \emph {et~al.}(2016)\citenamefont {Lee}, \citenamefont {Lee}, \citenamefont {Ryoo}, \citenamefont {Kang}, \citenamefont {Kim}, \citenamefont {Kim}, \citenamefont {Park}, \citenamefont {Park},\ and\ \citenamefont {Cheong}}]{2016_Lee_NanoLett}%
  \BibitemOpen
  \bibfield  {author} {\bibinfo {author} {\bibfnamefont {J.-U.}\ \bibnamefont {Lee}}, \bibinfo {author} {\bibfnamefont {S.}~\bibnamefont {Lee}}, \bibinfo {author} {\bibfnamefont {J.~H.}\ \bibnamefont {Ryoo}}, \bibinfo {author} {\bibfnamefont {S.}~\bibnamefont {Kang}}, \bibinfo {author} {\bibfnamefont {T.~Y.}\ \bibnamefont {Kim}}, \bibinfo {author} {\bibfnamefont {P.}~\bibnamefont {Kim}}, \bibinfo {author} {\bibfnamefont {C.-H.}\ \bibnamefont {Park}}, \bibinfo {author} {\bibfnamefont {J.-G.}\ \bibnamefont {Park}},\ and\ \bibinfo {author} {\bibfnamefont {H.}~\bibnamefont {Cheong}},\ }\bibfield  {title} {\bibinfo {title} {Ising-{{Type Magnetic Ordering}} in {{Atomically Thin FePS$_3$}}},\ }\href {https://doi.org/10.1021/acs.nanolett.6b03052} {\bibfield  {journal} {\bibinfo  {journal} {Nano Letters}\ }\textbf {\bibinfo {volume} {16}},\ \bibinfo {pages} {7433} (\bibinfo {year} {2016})}\BibitemShut {NoStop}%
\bibitem [{\citenamefont {Mak}\ and\ \citenamefont {Shan}(2016)}]{2016_Mak_NaturePhoton}%
  \BibitemOpen
  \bibfield  {author} {\bibinfo {author} {\bibfnamefont {K.~F.}\ \bibnamefont {Mak}}\ and\ \bibinfo {author} {\bibfnamefont {J.}~\bibnamefont {Shan}},\ }\bibfield  {title} {\bibinfo {title} {Photonics and optoelectronics of {{2D}} semiconductor transition metal dichalcogenides},\ }\href {https://doi.org/10.1038/nphoton.2015.282} {\bibfield  {journal} {\bibinfo  {journal} {Nature Photonics}\ }\textbf {\bibinfo {volume} {10}},\ \bibinfo {pages} {216} (\bibinfo {year} {2016})}\BibitemShut {NoStop}%
\bibitem [{\citenamefont {Va{\v n}o}\ \emph {et~al.}(2021)\citenamefont {Va{\v n}o}, \citenamefont {Amini}, \citenamefont {Ganguli}, \citenamefont {Chen}, \citenamefont {Lado}, \citenamefont {Kezilebieke},\ and\ \citenamefont {Liljeroth}}]{2021_Vano_Nature}%
  \BibitemOpen
  \bibfield  {author} {\bibinfo {author} {\bibfnamefont {V.}~\bibnamefont {Va{\v n}o}}, \bibinfo {author} {\bibfnamefont {M.}~\bibnamefont {Amini}}, \bibinfo {author} {\bibfnamefont {S.~C.}\ \bibnamefont {Ganguli}}, \bibinfo {author} {\bibfnamefont {G.}~\bibnamefont {Chen}}, \bibinfo {author} {\bibfnamefont {J.~L.}\ \bibnamefont {Lado}}, \bibinfo {author} {\bibfnamefont {S.}~\bibnamefont {Kezilebieke}},\ and\ \bibinfo {author} {\bibfnamefont {P.}~\bibnamefont {Liljeroth}},\ }\bibfield  {title} {\bibinfo {title} {Artificial heavy fermions in a van der {{Waals}} heterostructure},\ }\href {https://doi.org/10.1038/s41586-021-04021-0} {\bibfield  {journal} {\bibinfo  {journal} {Nature}\ }\textbf {\bibinfo {volume} {599}},\ \bibinfo {pages} {582} (\bibinfo {year} {2021})}\BibitemShut {NoStop}%
\bibitem [{\citenamefont {Mizukami}\ \emph {et~al.}(2011)\citenamefont {Mizukami}, \citenamefont {Shishido}, \citenamefont {Shibauchi}, \citenamefont {Shimozawa}, \citenamefont {Yasumoto}, \citenamefont {Watanabe}, \citenamefont {Yamashita}, \citenamefont {Ikeda}, \citenamefont {Terashima}, \citenamefont {Kontani},\ and\ \citenamefont {Matsuda}}]{2011_Mizukami_NaturePhys}%
  \BibitemOpen
  \bibfield  {author} {\bibinfo {author} {\bibfnamefont {Y.}~\bibnamefont {Mizukami}}, \bibinfo {author} {\bibfnamefont {H.}~\bibnamefont {Shishido}}, \bibinfo {author} {\bibfnamefont {T.}~\bibnamefont {Shibauchi}}, \bibinfo {author} {\bibfnamefont {M.}~\bibnamefont {Shimozawa}}, \bibinfo {author} {\bibfnamefont {S.}~\bibnamefont {Yasumoto}}, \bibinfo {author} {\bibfnamefont {D.}~\bibnamefont {Watanabe}}, \bibinfo {author} {\bibfnamefont {M.}~\bibnamefont {Yamashita}}, \bibinfo {author} {\bibfnamefont {H.}~\bibnamefont {Ikeda}}, \bibinfo {author} {\bibfnamefont {T.}~\bibnamefont {Terashima}}, \bibinfo {author} {\bibfnamefont {H.}~\bibnamefont {Kontani}},\ and\ \bibinfo {author} {\bibfnamefont {Y.}~\bibnamefont {Matsuda}},\ }\bibfield  {title} {\bibinfo {title} {Extremely strong-coupling superconductivity in artificial two-dimensional {{Kondo}} lattices},\ }\href {https://doi.org/10.1038/nphys2112} {\bibfield  {journal} {\bibinfo  {journal} {Nature Physics}\ }\textbf {\bibinfo {volume} {7}},\ \bibinfo {pages}
  {849} (\bibinfo {year} {2011})}\BibitemShut {NoStop}%
\bibitem [{\citenamefont {Kosterlitz}\ and\ \citenamefont {Thouless}(1978)}]{KOSTERLITZ1978371}%
  \BibitemOpen
  \bibfield  {author} {\bibinfo {author} {\bibfnamefont {J.}~\bibnamefont {Kosterlitz}}\ and\ \bibinfo {author} {\bibfnamefont {D.}~\bibnamefont {Thouless}},\ }\bibfield  {title} {\bibinfo {title} {Chapter 5 two-dimensional physics}\ }(\bibinfo  {publisher} {Elsevier},\ \bibinfo {year} {1978})\ pp.\ \bibinfo {pages} {371--433}\BibitemShut {NoStop}%
\bibitem [{\citenamefont {Gong}\ \emph {et~al.}(2017)\citenamefont {Gong}, \citenamefont {Li}, \citenamefont {Li}, \citenamefont {Ji}, \citenamefont {Stern}, \citenamefont {Xia}, \citenamefont {Cao}, \citenamefont {Bao}, \citenamefont {Wang}, \citenamefont {Wang}, \citenamefont {Qiu}, \citenamefont {Cava}, \citenamefont {Louie}, \citenamefont {Xia},\ and\ \citenamefont {Zhang}}]{2017_Gong_Nature}%
  \BibitemOpen
  \bibfield  {author} {\bibinfo {author} {\bibfnamefont {C.}~\bibnamefont {Gong}}, \bibinfo {author} {\bibfnamefont {L.}~\bibnamefont {Li}}, \bibinfo {author} {\bibfnamefont {Z.}~\bibnamefont {Li}}, \bibinfo {author} {\bibfnamefont {H.}~\bibnamefont {Ji}}, \bibinfo {author} {\bibfnamefont {A.}~\bibnamefont {Stern}}, \bibinfo {author} {\bibfnamefont {Y.}~\bibnamefont {Xia}}, \bibinfo {author} {\bibfnamefont {T.}~\bibnamefont {Cao}}, \bibinfo {author} {\bibfnamefont {W.}~\bibnamefont {Bao}}, \bibinfo {author} {\bibfnamefont {C.}~\bibnamefont {Wang}}, \bibinfo {author} {\bibfnamefont {Y.}~\bibnamefont {Wang}}, \bibinfo {author} {\bibfnamefont {Z.~Q.}\ \bibnamefont {Qiu}}, \bibinfo {author} {\bibfnamefont {R.~J.}\ \bibnamefont {Cava}}, \bibinfo {author} {\bibfnamefont {S.~G.}\ \bibnamefont {Louie}}, \bibinfo {author} {\bibfnamefont {J.}~\bibnamefont {Xia}},\ and\ \bibinfo {author} {\bibfnamefont {X.}~\bibnamefont {Zhang}},\ }\bibfield  {title} {\bibinfo {title} {Discovery of intrinsic ferromagnetism in
  two-dimensional van der {{Waals}} crystals},\ }\href {https://doi.org/10.1038/nature22060} {\bibfield  {journal} {\bibinfo  {journal} {Nature}\ }\textbf {\bibinfo {volume} {546}},\ \bibinfo {pages} {265} (\bibinfo {year} {2017})}\BibitemShut {NoStop}%
\bibitem [{\citenamefont {Huang}\ \emph {et~al.}(2017)\citenamefont {Huang}, \citenamefont {Clark}, \citenamefont {{Navarro-Moratalla}}, \citenamefont {Klein}, \citenamefont {Cheng}, \citenamefont {Seyler}, \citenamefont {Zhong}, \citenamefont {Schmidgall}, \citenamefont {McGuire}, \citenamefont {Cobden}, \citenamefont {Yao}, \citenamefont {Xiao}, \citenamefont {{Jarillo-Herrero}},\ and\ \citenamefont {Xu}}]{2017_Huang_Nature}%
  \BibitemOpen
  \bibfield  {author} {\bibinfo {author} {\bibfnamefont {B.}~\bibnamefont {Huang}}, \bibinfo {author} {\bibfnamefont {G.}~\bibnamefont {Clark}}, \bibinfo {author} {\bibfnamefont {E.}~\bibnamefont {{Navarro-Moratalla}}}, \bibinfo {author} {\bibfnamefont {D.~R.}\ \bibnamefont {Klein}}, \bibinfo {author} {\bibfnamefont {R.}~\bibnamefont {Cheng}}, \bibinfo {author} {\bibfnamefont {K.~L.}\ \bibnamefont {Seyler}}, \bibinfo {author} {\bibfnamefont {D.}~\bibnamefont {Zhong}}, \bibinfo {author} {\bibfnamefont {E.}~\bibnamefont {Schmidgall}}, \bibinfo {author} {\bibfnamefont {M.~A.}\ \bibnamefont {McGuire}}, \bibinfo {author} {\bibfnamefont {D.~H.}\ \bibnamefont {Cobden}}, \bibinfo {author} {\bibfnamefont {W.}~\bibnamefont {Yao}}, \bibinfo {author} {\bibfnamefont {D.}~\bibnamefont {Xiao}}, \bibinfo {author} {\bibfnamefont {P.}~\bibnamefont {{Jarillo-Herrero}}},\ and\ \bibinfo {author} {\bibfnamefont {X.}~\bibnamefont {Xu}},\ }\bibfield  {title} {\bibinfo {title} {Layer-dependent ferromagnetism in a van der {{Waals}}
  crystal down to the monolayer limit},\ }\href {https://doi.org/10.1038/nature22391} {\bibfield  {journal} {\bibinfo  {journal} {Nature}\ }\textbf {\bibinfo {volume} {546}},\ \bibinfo {pages} {270} (\bibinfo {year} {2017})}\BibitemShut {NoStop}%
\bibitem [{\citenamefont {Burch}\ \emph {et~al.}(2018)\citenamefont {Burch}, \citenamefont {Mandrus},\ and\ \citenamefont {Park}}]{2018_Burch_Nature}%
  \BibitemOpen
  \bibfield  {author} {\bibinfo {author} {\bibfnamefont {K.~S.}\ \bibnamefont {Burch}}, \bibinfo {author} {\bibfnamefont {D.}~\bibnamefont {Mandrus}},\ and\ \bibinfo {author} {\bibfnamefont {J.-G.}\ \bibnamefont {Park}},\ }\bibfield  {title} {\bibinfo {title} {Magnetism in two-dimensional van der {{Waals}} materials},\ }\href {https://doi.org/10.1038/s41586-018-0631-z} {\bibfield  {journal} {\bibinfo  {journal} {Nature}\ }\textbf {\bibinfo {volume} {563}},\ \bibinfo {pages} {47} (\bibinfo {year} {2018})}\BibitemShut {NoStop}%
\bibitem [{\citenamefont {Ziebel}\ \emph {et~al.}(2024)\citenamefont {Ziebel}, \citenamefont {Feuer}, \citenamefont {Cox}, \citenamefont {Zhu}, \citenamefont {Dean},\ and\ \citenamefont {Roy}}]{2024_Ziebel_NanoLett}%
  \BibitemOpen
  \bibfield  {author} {\bibinfo {author} {\bibfnamefont {M.~E.}\ \bibnamefont {Ziebel}}, \bibinfo {author} {\bibfnamefont {M.~L.}\ \bibnamefont {Feuer}}, \bibinfo {author} {\bibfnamefont {J.}~\bibnamefont {Cox}}, \bibinfo {author} {\bibfnamefont {X.}~\bibnamefont {Zhu}}, \bibinfo {author} {\bibfnamefont {C.~R.}\ \bibnamefont {Dean}},\ and\ \bibinfo {author} {\bibfnamefont {X.}~\bibnamefont {Roy}},\ }\bibfield  {title} {\bibinfo {title} {{{CrSBr}}: {{An Air-Stable}}, {{Two-Dimensional Magnetic Semiconductor}}},\ }\href {https://doi.org/10.1021/acs.nanolett.4c00624} {\bibfield  {journal} {\bibinfo  {journal} {Nano Letters}\ }\textbf {\bibinfo {volume} {24}},\ \bibinfo {pages} {4319} (\bibinfo {year} {2024})}\BibitemShut {NoStop}%
\bibitem [{\citenamefont {Thomas}\ \emph {et~al.}(2025)\citenamefont {Thomas}, \citenamefont {Llacsahuanga}, \citenamefont {Simeth}, \citenamefont {Kengle}, \citenamefont {Orlandi}, \citenamefont {Khalyavin}, \citenamefont {Manuel}, \citenamefont {Ronning}, \citenamefont {Bauer}, \citenamefont {Thompson}, \citenamefont {Zhu}, \citenamefont {Scheie}, \citenamefont {Chen},\ and\ \citenamefont {Rosa}}]{2025_Thomas_}%
  \BibitemOpen
  \bibfield  {author} {\bibinfo {author} {\bibfnamefont {S.~M.}\ \bibnamefont {Thomas}}, \bibinfo {author} {\bibfnamefont {A.~E.}\ \bibnamefont {Llacsahuanga}}, \bibinfo {author} {\bibfnamefont {W.}~\bibnamefont {Simeth}}, \bibinfo {author} {\bibfnamefont {C.~S.}\ \bibnamefont {Kengle}}, \bibinfo {author} {\bibfnamefont {F.}~\bibnamefont {Orlandi}}, \bibinfo {author} {\bibfnamefont {D.}~\bibnamefont {Khalyavin}}, \bibinfo {author} {\bibfnamefont {P.}~\bibnamefont {Manuel}}, \bibinfo {author} {\bibfnamefont {F.}~\bibnamefont {Ronning}}, \bibinfo {author} {\bibfnamefont {E.~D.}\ \bibnamefont {Bauer}}, \bibinfo {author} {\bibfnamefont {J.~D.}\ \bibnamefont {Thompson}}, \bibinfo {author} {\bibfnamefont {J.-X.}\ \bibnamefont {Zhu}}, \bibinfo {author} {\bibfnamefont {A.~O.}\ \bibnamefont {Scheie}}, \bibinfo {author} {\bibfnamefont {Y.~P.}\ \bibnamefont {Chen}},\ and\ \bibinfo {author} {\bibfnamefont {P.~F.~S.}\ \bibnamefont {Rosa}},\ }\href {https://doi.org/10.48550/arXiv.2506.15667} {\bibinfo {title} {Enhanced
  two-dimensional ferromagnetism in van der {{Waals}} $\beta$-{{U}}{{Te}}$_3$ monolayers}} (\bibinfo {year} {2025}),\ \Eprint {https://arxiv.org/abs/2506.15667} {arXiv:2506.15667 [cond-mat]} \BibitemShut {NoStop}%
\bibitem [{\citenamefont {Novoselov}\ \emph {et~al.}(2016)\citenamefont {Novoselov}, \citenamefont {Mishchenko}, \citenamefont {Carvalho},\ and\ \citenamefont {Castro~Neto}}]{2016_Novoselov_Science}%
  \BibitemOpen
  \bibfield  {author} {\bibinfo {author} {\bibfnamefont {K.~S.}\ \bibnamefont {Novoselov}}, \bibinfo {author} {\bibfnamefont {A.}~\bibnamefont {Mishchenko}}, \bibinfo {author} {\bibfnamefont {A.}~\bibnamefont {Carvalho}},\ and\ \bibinfo {author} {\bibfnamefont {A.~H.}\ \bibnamefont {Castro~Neto}},\ }\bibfield  {title} {\bibinfo {title} {{{2D}} materials and van der {{Waals}} heterostructures},\ }\href {https://doi.org/10.1126/science.aac9439} {\bibfield  {journal} {\bibinfo  {journal} {Science}\ }\textbf {\bibinfo {volume} {353}},\ \bibinfo {pages} {aac9439} (\bibinfo {year} {2016})}\BibitemShut {NoStop}%
\bibitem [{\citenamefont {Liu}\ \emph {et~al.}(2016)\citenamefont {Liu}, \citenamefont {Weiss}, \citenamefont {Duan}, \citenamefont {Cheng}, \citenamefont {Huang},\ and\ \citenamefont {Duan}}]{2016_Liu_NatRevMater}%
  \BibitemOpen
  \bibfield  {author} {\bibinfo {author} {\bibfnamefont {Y.}~\bibnamefont {Liu}}, \bibinfo {author} {\bibfnamefont {N.~O.}\ \bibnamefont {Weiss}}, \bibinfo {author} {\bibfnamefont {X.}~\bibnamefont {Duan}}, \bibinfo {author} {\bibfnamefont {H.-C.}\ \bibnamefont {Cheng}}, \bibinfo {author} {\bibfnamefont {Y.}~\bibnamefont {Huang}},\ and\ \bibinfo {author} {\bibfnamefont {X.}~\bibnamefont {Duan}},\ }\bibfield  {title} {\bibinfo {title} {Van der {{Waals}} heterostructures and devices},\ }\href {https://doi.org/10.1038/natrevmats.2016.42} {\bibfield  {journal} {\bibinfo  {journal} {Nature Reviews Materials}\ }\textbf {\bibinfo {volume} {1}},\ \bibinfo {pages} {16042} (\bibinfo {year} {2016})}\BibitemShut {NoStop}%
\bibitem [{\citenamefont {Jariwala}\ \emph {et~al.}(2017)\citenamefont {Jariwala}, \citenamefont {Davoyan}, \citenamefont {Wong},\ and\ \citenamefont {Atwater}}]{2017_Jariwala_ACSPhotonics}%
  \BibitemOpen
  \bibfield  {author} {\bibinfo {author} {\bibfnamefont {D.}~\bibnamefont {Jariwala}}, \bibinfo {author} {\bibfnamefont {A.~R.}\ \bibnamefont {Davoyan}}, \bibinfo {author} {\bibfnamefont {J.}~\bibnamefont {Wong}},\ and\ \bibinfo {author} {\bibfnamefont {H.~A.}\ \bibnamefont {Atwater}},\ }\bibfield  {title} {\bibinfo {title} {Van der {{Waals Materials}} for {{Atomically-Thin Photovoltaics}}: {{Promise}} and {{Outlook}}},\ }\href {https://doi.org/10.1021/acsphotonics.7b01103} {\bibfield  {journal} {\bibinfo  {journal} {ACS Photonics}\ }\textbf {\bibinfo {volume} {4}},\ \bibinfo {pages} {2962} (\bibinfo {year} {2017})}\BibitemShut {NoStop}%
\bibitem [{\citenamefont {Zotev}\ \emph {et~al.}(2025)\citenamefont {Zotev}, \citenamefont {Bouteyre}, \citenamefont {Wang}, \citenamefont {Randerson}, \citenamefont {Hu}, \citenamefont {Sortino}, \citenamefont {Wang}, \citenamefont {Shegai}, \citenamefont {Gong}, \citenamefont {Tittl}, \citenamefont {Aharonovich},\ and\ \citenamefont {Tartakovskii}}]{2025_Zotev_NatPhoton}%
  \BibitemOpen
  \bibfield  {author} {\bibinfo {author} {\bibfnamefont {P.~G.}\ \bibnamefont {Zotev}}, \bibinfo {author} {\bibfnamefont {P.}~\bibnamefont {Bouteyre}}, \bibinfo {author} {\bibfnamefont {Y.}~\bibnamefont {Wang}}, \bibinfo {author} {\bibfnamefont {S.~A.}\ \bibnamefont {Randerson}}, \bibinfo {author} {\bibfnamefont {X.}~\bibnamefont {Hu}}, \bibinfo {author} {\bibfnamefont {L.}~\bibnamefont {Sortino}}, \bibinfo {author} {\bibfnamefont {Y.}~\bibnamefont {Wang}}, \bibinfo {author} {\bibfnamefont {T.}~\bibnamefont {Shegai}}, \bibinfo {author} {\bibfnamefont {S.-H.}\ \bibnamefont {Gong}}, \bibinfo {author} {\bibfnamefont {A.}~\bibnamefont {Tittl}}, \bibinfo {author} {\bibfnamefont {I.}~\bibnamefont {Aharonovich}},\ and\ \bibinfo {author} {\bibfnamefont {A.~I.}\ \bibnamefont {Tartakovskii}},\ }\bibfield  {title} {\bibinfo {title} {Nanophotonics with multilayer van der {{Waals}} materials},\ }\href {https://doi.org/10.1038/s41566-025-01717-x} {\bibfield  {journal} {\bibinfo  {journal} {Nature Photonics}\ }\textbf
  {\bibinfo {volume} {19}},\ \bibinfo {pages} {788} (\bibinfo {year} {2025})}\BibitemShut {NoStop}%
\bibitem [{\citenamefont {Jang}\ \emph {et~al.}(2022)\citenamefont {Jang}, \citenamefont {Lee}, \citenamefont {Zhu},\ and\ \citenamefont {Shim}}]{Jang2022}%
  \BibitemOpen
  \bibfield  {author} {\bibinfo {author} {\bibfnamefont {B.~G.}\ \bibnamefont {Jang}}, \bibinfo {author} {\bibfnamefont {C.}~\bibnamefont {Lee}}, \bibinfo {author} {\bibfnamefont {J.-X.}\ \bibnamefont {Zhu}},\ and\ \bibinfo {author} {\bibfnamefont {J.~H.}\ \bibnamefont {Shim}},\ }\bibfield  {title} {\bibinfo {title} {Exploring two-dimensional van der {{Waals}} heavy-fermion material: {{Data}} mining theoretical approach},\ }\href {https://doi.org/10.1038/s41699-022-00357-x} {\bibfield  {journal} {\bibinfo  {journal} {npj 2D Materials and Applications}\ }\textbf {\bibinfo {volume} {6}},\ \bibinfo {pages} {80} (\bibinfo {year} {2022})}\BibitemShut {NoStop}%
\bibitem [{\citenamefont {Posey}\ \emph {et~al.}(2024)\citenamefont {Posey}, \citenamefont {Turkel}, \citenamefont {Rezaee}, \citenamefont {Devarakonda}, \citenamefont {Kundu}, \citenamefont {Ong}, \citenamefont {Thinel}, \citenamefont {Chica}, \citenamefont {Vitalone}, \citenamefont {Jing} \emph {et~al.}}]{posey2024CeSiI}%
  \BibitemOpen
  \bibfield  {author} {\bibinfo {author} {\bibfnamefont {V.~A.}\ \bibnamefont {Posey}}, \bibinfo {author} {\bibfnamefont {S.}~\bibnamefont {Turkel}}, \bibinfo {author} {\bibfnamefont {M.}~\bibnamefont {Rezaee}}, \bibinfo {author} {\bibfnamefont {A.}~\bibnamefont {Devarakonda}}, \bibinfo {author} {\bibfnamefont {A.~K.}\ \bibnamefont {Kundu}}, \bibinfo {author} {\bibfnamefont {C.~S.}\ \bibnamefont {Ong}}, \bibinfo {author} {\bibfnamefont {M.}~\bibnamefont {Thinel}}, \bibinfo {author} {\bibfnamefont {D.~G.}\ \bibnamefont {Chica}}, \bibinfo {author} {\bibfnamefont {R.~A.}\ \bibnamefont {Vitalone}}, \bibinfo {author} {\bibfnamefont {R.}~\bibnamefont {Jing}}, \emph {et~al.},\ }\bibfield  {title} {\bibinfo {title} {Two-dimensional heavy fermions in the van der {Waals} metal {CeSiI}},\ }\href {https://doi.org/10.1038/s41586-023-06868-x} {\bibfield  {journal} {\bibinfo  {journal} {Nature}\ }\textbf {\bibinfo {volume} {625}},\ \bibinfo {pages} {483} (\bibinfo {year} {2024})}\BibitemShut {NoStop}%
\bibitem [{\citenamefont {Shi}\ \emph {et~al.}(2026)\citenamefont {Shi}, \citenamefont {Li}, \citenamefont {Dong}, \citenamefont {Yang}, \citenamefont {Ma}, \citenamefont {Tian}, \citenamefont {Wang}, \citenamefont {Sun}, \citenamefont {Uwatoko}, \citenamefont {Yang}, \citenamefont {Wang}, \citenamefont {Lei},\ and\ \citenamefont {Cheng}}]{2026_Shi_}%
  \BibitemOpen
  \bibfield  {author} {\bibinfo {author} {\bibfnamefont {T.}~\bibnamefont {Shi}}, \bibinfo {author} {\bibfnamefont {W.}~\bibnamefont {Li}}, \bibinfo {author} {\bibfnamefont {Q.}~\bibnamefont {Dong}}, \bibinfo {author} {\bibfnamefont {P.}~\bibnamefont {Yang}}, \bibinfo {author} {\bibfnamefont {H.}~\bibnamefont {Ma}}, \bibinfo {author} {\bibfnamefont {Z.}~\bibnamefont {Tian}}, \bibinfo {author} {\bibfnamefont {N.}~\bibnamefont {Wang}}, \bibinfo {author} {\bibfnamefont {J.}~\bibnamefont {Sun}}, \bibinfo {author} {\bibfnamefont {Y.}~\bibnamefont {Uwatoko}}, \bibinfo {author} {\bibfnamefont {Y.-f.}\ \bibnamefont {Yang}}, \bibinfo {author} {\bibfnamefont {B.}~\bibnamefont {Wang}}, \bibinfo {author} {\bibfnamefont {H.}~\bibnamefont {Lei}},\ and\ \bibinfo {author} {\bibfnamefont {J.}~\bibnamefont {Cheng}},\ }\href {https://doi.org/10.48550/arXiv.2601.18476} {\bibinfo {title} {Superconductivity under pressure in the two-dimensional van der {{Waals}} heavy-fermion metal {{CeSiI}}}} (\bibinfo {year} {2026}),\ \Eprint
  {https://arxiv.org/abs/2601.18476} {arXiv:2601.18476 [cond-mat]} \BibitemShut {NoStop}%
\bibitem [{\citenamefont {Kallin}\ and\ \citenamefont {Berlinsky}(2016)}]{2016_Kallin_RepProgPhys}%
  \BibitemOpen
  \bibfield  {author} {\bibinfo {author} {\bibfnamefont {C.}~\bibnamefont {Kallin}}\ and\ \bibinfo {author} {\bibfnamefont {J.}~\bibnamefont {Berlinsky}},\ }\bibfield  {title} {\bibinfo {title} {Chiral superconductors},\ }\href {https://doi.org/10.1088/0034-4885/79/5/054502} {\bibfield  {journal} {\bibinfo  {journal} {Reports on Progress in Physics}\ }\textbf {\bibinfo {volume} {79}},\ \bibinfo {pages} {054502} (\bibinfo {year} {2016})}\BibitemShut {NoStop}%
\bibitem [{\citenamefont {Sarma}\ \emph {et~al.}(2015)\citenamefont {Sarma}, \citenamefont {Freedman},\ and\ \citenamefont {Nayak}}]{2015_Sarma_npjQuantumInf}%
  \BibitemOpen
  \bibfield  {author} {\bibinfo {author} {\bibfnamefont {S.~D.}\ \bibnamefont {Sarma}}, \bibinfo {author} {\bibfnamefont {M.}~\bibnamefont {Freedman}},\ and\ \bibinfo {author} {\bibfnamefont {C.}~\bibnamefont {Nayak}},\ }\bibfield  {title} {\bibinfo {title} {Majorana zero modes and topological quantum computation},\ }\href {https://doi.org/10.1038/npjqi.2015.1} {\bibfield  {journal} {\bibinfo  {journal} {npj Quantum Information}\ }\textbf {\bibinfo {volume} {1}},\ \bibinfo {pages} {15001} (\bibinfo {year} {2015})}\BibitemShut {NoStop}%
\bibitem [{\citenamefont {Karzig}\ \emph {et~al.}(2017)\citenamefont {Karzig}, \citenamefont {Knapp}, \citenamefont {Lutchyn}, \citenamefont {Bonderson}, \citenamefont {Hastings}, \citenamefont {Nayak}, \citenamefont {Alicea}, \citenamefont {Flensberg}, \citenamefont {Plugge}, \citenamefont {Oreg}, \citenamefont {Marcus},\ and\ \citenamefont {Freedman}}]{2017_Karzig_PhysRevB}%
  \BibitemOpen
  \bibfield  {author} {\bibinfo {author} {\bibfnamefont {T.}~\bibnamefont {Karzig}}, \bibinfo {author} {\bibfnamefont {C.}~\bibnamefont {Knapp}}, \bibinfo {author} {\bibfnamefont {R.~M.}\ \bibnamefont {Lutchyn}}, \bibinfo {author} {\bibfnamefont {P.}~\bibnamefont {Bonderson}}, \bibinfo {author} {\bibfnamefont {M.~B.}\ \bibnamefont {Hastings}}, \bibinfo {author} {\bibfnamefont {C.}~\bibnamefont {Nayak}}, \bibinfo {author} {\bibfnamefont {J.}~\bibnamefont {Alicea}}, \bibinfo {author} {\bibfnamefont {K.}~\bibnamefont {Flensberg}}, \bibinfo {author} {\bibfnamefont {S.}~\bibnamefont {Plugge}}, \bibinfo {author} {\bibfnamefont {Y.}~\bibnamefont {Oreg}}, \bibinfo {author} {\bibfnamefont {C.~M.}\ \bibnamefont {Marcus}},\ and\ \bibinfo {author} {\bibfnamefont {M.~H.}\ \bibnamefont {Freedman}},\ }\bibfield  {title} {\bibinfo {title} {Scalable designs for quasiparticle-poisoning-protected topological quantum computation with {{Majorana}} zero modes},\ }\href {https://doi.org/10.1103/PhysRevB.95.235305} {\bibfield
  {journal} {\bibinfo  {journal} {Physical Review B}\ }\textbf {\bibinfo {volume} {95}},\ \bibinfo {pages} {235305} (\bibinfo {year} {2017})}\BibitemShut {NoStop}%
\bibitem [{\citenamefont {Sato}\ and\ \citenamefont {Ando}(2017)}]{2017_Sato_RepProgPhysa}%
  \BibitemOpen
  \bibfield  {author} {\bibinfo {author} {\bibfnamefont {M.}~\bibnamefont {Sato}}\ and\ \bibinfo {author} {\bibfnamefont {Y.}~\bibnamefont {Ando}},\ }\bibfield  {title} {\bibinfo {title} {Topological superconductors: A review},\ }\href {https://doi.org/10.1088/1361-6633/aa6ac7} {\bibfield  {journal} {\bibinfo  {journal} {Reports on Progress in Physics}\ }\textbf {\bibinfo {volume} {80}},\ \bibinfo {pages} {076501} (\bibinfo {year} {2017})}\BibitemShut {NoStop}%
\bibitem [{\citenamefont {Yadav}\ \emph {et~al.}(2023)\citenamefont {Yadav}, \citenamefont {Acosta}, \citenamefont {Dalpian},\ and\ \citenamefont {Malyi}}]{2023_Yadav_Matter}%
  \BibitemOpen
  \bibfield  {author} {\bibinfo {author} {\bibfnamefont {A.}~\bibnamefont {Yadav}}, \bibinfo {author} {\bibfnamefont {C.~M.}\ \bibnamefont {Acosta}}, \bibinfo {author} {\bibfnamefont {G.~M.}\ \bibnamefont {Dalpian}},\ and\ \bibinfo {author} {\bibfnamefont {O.~I.}\ \bibnamefont {Malyi}},\ }\bibfield  {title} {\bibinfo {title} {First-principles investigations of {{2D}} materials: {{Challenges}} and best practices},\ }\href {https://doi.org/10.1016/j.matt.2023.05.019} {\bibfield  {journal} {\bibinfo  {journal} {Matter}\ }\textbf {\bibinfo {volume} {6}},\ \bibinfo {pages} {2711} (\bibinfo {year} {2023})}\BibitemShut {NoStop}%
\bibitem [{\citenamefont {Stoudenmire}\ and\ \citenamefont {White}(2012)}]{Stoudenmire2012_DMRG}%
  \BibitemOpen
  \bibfield  {author} {\bibinfo {author} {\bibfnamefont {E.}~\bibnamefont {Stoudenmire}}\ and\ \bibinfo {author} {\bibfnamefont {S.~R.}\ \bibnamefont {White}},\ }\bibfield  {title} {\bibinfo {title} {Studying two-dimensional systems with the density matrix renormalization group},\ }\href {https://doi.org/https://doi.org/10.1146/annurev-conmatphys-020911-125018} {\bibfield  {journal} {\bibinfo  {journal} {Annual Review of Condensed Matter Physics}\ }\textbf {\bibinfo {volume} {3}},\ \bibinfo {pages} {111} (\bibinfo {year} {2012})}\BibitemShut {NoStop}%
\bibitem [{\citenamefont {Pan}\ and\ \citenamefont {Assaad}(2025)}]{pan2025quantum}%
  \BibitemOpen
  \bibfield  {author} {\bibinfo {author} {\bibfnamefont {G.}~\bibnamefont {Pan}}\ and\ \bibinfo {author} {\bibfnamefont {F.~F.}\ \bibnamefont {Assaad}},\ }\bibfield  {title} {\bibinfo {title} {Quantum monte carlo studies of {U}(1) lattice gauge models of kondo breakdown},\ }\bibfield  {journal} {\bibinfo  {journal} {arXiv preprint arXiv:2512.17801}\ }\href {https://doi.org/10.48550/arXiv.2512.17801} {10.48550/arXiv.2512.17801} (\bibinfo {year} {2025})\BibitemShut {NoStop}%
\bibitem [{\citenamefont {Monthoux}\ \emph {et~al.}(2007)\citenamefont {Monthoux}, \citenamefont {Pines},\ and\ \citenamefont {Lonzarich}}]{2007_Monthoux_Naturea}%
  \BibitemOpen
  \bibfield  {author} {\bibinfo {author} {\bibfnamefont {P.}~\bibnamefont {Monthoux}}, \bibinfo {author} {\bibfnamefont {D.}~\bibnamefont {Pines}},\ and\ \bibinfo {author} {\bibfnamefont {G.~G.}\ \bibnamefont {Lonzarich}},\ }\bibfield  {title} {\bibinfo {title} {Superconductivity without phonons},\ }\href {https://doi.org/10.1038/nature06480} {\bibfield  {journal} {\bibinfo  {journal} {Nature}\ }\textbf {\bibinfo {volume} {450}},\ \bibinfo {pages} {1177} (\bibinfo {year} {2007})}\BibitemShut {NoStop}%
\bibitem [{\citenamefont {Turkel}\ \emph {et~al.}(2025)\citenamefont {Turkel}, \citenamefont {Posey}, \citenamefont {Ong}, \citenamefont {Ghosh}, \citenamefont {Huang}, \citenamefont {Kundu}, \citenamefont {Vescovo}, \citenamefont {Chica}, \citenamefont {Thunstr{\"o}m}, \citenamefont {Eriksson}, \citenamefont {Simeth}, \citenamefont {Scheie}, \citenamefont {Rubio}, \citenamefont {Millis}, \citenamefont {Roy},\ and\ \citenamefont {Pasupathy}}]{2025_Turkel_NatPhysa}%
  \BibitemOpen
  \bibfield  {author} {\bibinfo {author} {\bibfnamefont {S.}~\bibnamefont {Turkel}}, \bibinfo {author} {\bibfnamefont {V.~A.}\ \bibnamefont {Posey}}, \bibinfo {author} {\bibfnamefont {C.~S.}\ \bibnamefont {Ong}}, \bibinfo {author} {\bibfnamefont {S.}~\bibnamefont {Ghosh}}, \bibinfo {author} {\bibfnamefont {X.}~\bibnamefont {Huang}}, \bibinfo {author} {\bibfnamefont {A.~K.}\ \bibnamefont {Kundu}}, \bibinfo {author} {\bibfnamefont {E.}~\bibnamefont {Vescovo}}, \bibinfo {author} {\bibfnamefont {D.~G.}\ \bibnamefont {Chica}}, \bibinfo {author} {\bibfnamefont {P.}~\bibnamefont {Thunstr{\"o}m}}, \bibinfo {author} {\bibfnamefont {O.}~\bibnamefont {Eriksson}}, \bibinfo {author} {\bibfnamefont {W.}~\bibnamefont {Simeth}}, \bibinfo {author} {\bibfnamefont {A.}~\bibnamefont {Scheie}}, \bibinfo {author} {\bibfnamefont {A.}~\bibnamefont {Rubio}}, \bibinfo {author} {\bibfnamefont {A.~J.}\ \bibnamefont {Millis}}, \bibinfo {author} {\bibfnamefont {X.}~\bibnamefont {Roy}},\ and\ \bibinfo {author} {\bibfnamefont {A.~N.}\
  \bibnamefont {Pasupathy}},\ }\bibfield  {title} {\bibinfo {title} {Nodal hybridization in a two-dimensional heavy-fermion material},\ }\href {https://doi.org/10.1038/s41567-025-03060-y} {\bibfield  {journal} {\bibinfo  {journal} {Nature Physics}\ }\textbf {\bibinfo {volume} {21}},\ \bibinfo {pages} {1949} (\bibinfo {year} {2025})}\BibitemShut {NoStop}%
\bibitem [{\citenamefont {Okuma}(2021)}]{2021_Okuma_PhysRevMater}%
  \BibitemOpen
  \bibfield  {author} {\bibinfo {author} {\bibfnamefont {R.}~\bibnamefont {Okuma}},\ }\bibfield  {title} {\bibinfo {title} {Magnetic frustration in a van der {{Waals}} metal {{CeSiI}}},\ }\bibfield  {journal} {\bibinfo  {journal} {Physical Review Materials}\ }\textbf {\bibinfo {volume} {5}},\ \href {https://doi.org/10.1103/PhysRevMaterials.5.L121401} {10.1103/PhysRevMaterials.5.L121401} (\bibinfo {year} {2021})\BibitemShut {NoStop}%
\bibitem [{\citenamefont {Squires}(2012)}]{Squires}%
  \BibitemOpen
  \bibfield  {author} {\bibinfo {author} {\bibfnamefont {G.~L.}\ \bibnamefont {Squires}},\ }\href@noop {} {\emph {\bibinfo {title} {Introduction to the Theory of Thermal Neutron Scattering}}},\ \bibinfo {edition} {3rd}\ ed.\ (\bibinfo  {publisher} {Cambridge University Press},\ \bibinfo {address} {Cambridge, UK},\ \bibinfo {year} {2012})\BibitemShut {NoStop}%
\bibitem [{\citenamefont {Ehlers}\ \emph {et~al.}(2011)\citenamefont {Ehlers}, \citenamefont {Podlesnyak}, \citenamefont {Niedziela}, \citenamefont {Iverson},\ and\ \citenamefont {Sokol}}]{CNCS}%
  \BibitemOpen
  \bibfield  {author} {\bibinfo {author} {\bibfnamefont {G.}~\bibnamefont {Ehlers}}, \bibinfo {author} {\bibfnamefont {A.~A.}\ \bibnamefont {Podlesnyak}}, \bibinfo {author} {\bibfnamefont {J.~L.}\ \bibnamefont {Niedziela}}, \bibinfo {author} {\bibfnamefont {E.~B.}\ \bibnamefont {Iverson}},\ and\ \bibinfo {author} {\bibfnamefont {P.~E.}\ \bibnamefont {Sokol}},\ }\bibfield  {title} {\bibinfo {title} {The new cold neutron chopper spectrometer at the spallation neutron source: Design and performance},\ }\href {https://doi.org/10.1063/1.3626935} {\bibfield  {journal} {\bibinfo  {journal} {Review of Scientific Instruments}\ }\textbf {\bibinfo {volume} {82}},\ \bibinfo {pages} {085108} (\bibinfo {year} {2011})}\BibitemShut {NoStop}%
\bibitem [{\citenamefont {Mason}\ \emph {et~al.}(2006)\citenamefont {Mason}, \citenamefont {Abernathy}, \citenamefont {Anderson}, \citenamefont {Ankner}, \citenamefont {Egami}, \citenamefont {Ehlers}, \citenamefont {Ekkebus}, \citenamefont {Granroth}, \citenamefont {Hagen}, \citenamefont {Herwig}, \citenamefont {Hodges}, \citenamefont {Hoffmann}, \citenamefont {Horak}, \citenamefont {Horton}, \citenamefont {Klose}, \citenamefont {Larese}, \citenamefont {Mesecar}, \citenamefont {Myles}, \citenamefont {Neuefeind}, \citenamefont {Ohl}, \citenamefont {Tulk}, \citenamefont {Wang},\ and\ \citenamefont {Zhao}}]{mason2006spallation}%
  \BibitemOpen
  \bibfield  {author} {\bibinfo {author} {\bibfnamefont {T.~E.}\ \bibnamefont {Mason}}, \bibinfo {author} {\bibfnamefont {D.}~\bibnamefont {Abernathy}}, \bibinfo {author} {\bibfnamefont {I.}~\bibnamefont {Anderson}}, \bibinfo {author} {\bibfnamefont {J.}~\bibnamefont {Ankner}}, \bibinfo {author} {\bibfnamefont {T.}~\bibnamefont {Egami}}, \bibinfo {author} {\bibfnamefont {G.}~\bibnamefont {Ehlers}}, \bibinfo {author} {\bibfnamefont {A.}~\bibnamefont {Ekkebus}}, \bibinfo {author} {\bibfnamefont {G.}~\bibnamefont {Granroth}}, \bibinfo {author} {\bibfnamefont {M.}~\bibnamefont {Hagen}}, \bibinfo {author} {\bibfnamefont {K.}~\bibnamefont {Herwig}}, \bibinfo {author} {\bibfnamefont {J.}~\bibnamefont {Hodges}}, \bibinfo {author} {\bibfnamefont {C.}~\bibnamefont {Hoffmann}}, \bibinfo {author} {\bibfnamefont {C.}~\bibnamefont {Horak}}, \bibinfo {author} {\bibfnamefont {L.}~\bibnamefont {Horton}}, \bibinfo {author} {\bibfnamefont {F.}~\bibnamefont {Klose}}, \bibinfo {author} {\bibfnamefont {J.}~\bibnamefont {Larese}},
  \bibinfo {author} {\bibfnamefont {A.}~\bibnamefont {Mesecar}}, \bibinfo {author} {\bibfnamefont {D.}~\bibnamefont {Myles}}, \bibinfo {author} {\bibfnamefont {J.}~\bibnamefont {Neuefeind}}, \bibinfo {author} {\bibfnamefont {M.}~\bibnamefont {Ohl}}, \bibinfo {author} {\bibfnamefont {C.}~\bibnamefont {Tulk}}, \bibinfo {author} {\bibfnamefont {X.-L.}\ \bibnamefont {Wang}},\ and\ \bibinfo {author} {\bibfnamefont {J.}~\bibnamefont {Zhao}},\ }\bibfield  {title} {\bibinfo {title} {{{The Spallation Neutron Source in Oak Ridge}}: {A} powerful tool for materials research},\ }\href {https://doi.org/10.1016/j.physb.2006.05.281} {\bibfield  {journal} {\bibinfo  {journal} {Physica B: Condensed Matter}\ }\textbf {\bibinfo {volume} {385}},\ \bibinfo {pages} {955} (\bibinfo {year} {2006})}\BibitemShut {NoStop}%
\bibitem [{\citenamefont {Dvorak}\ \emph {et~al.}(2016)\citenamefont {Dvorak}, \citenamefont {Jarrige}, \citenamefont {Bisogni}, \citenamefont {Coburn},\ and\ \citenamefont {Leonhardt}}]{2016_Dvorak_RevSciInstrum}%
  \BibitemOpen
  \bibfield  {author} {\bibinfo {author} {\bibfnamefont {J.}~\bibnamefont {Dvorak}}, \bibinfo {author} {\bibfnamefont {I.}~\bibnamefont {Jarrige}}, \bibinfo {author} {\bibfnamefont {V.}~\bibnamefont {Bisogni}}, \bibinfo {author} {\bibfnamefont {S.}~\bibnamefont {Coburn}},\ and\ \bibinfo {author} {\bibfnamefont {W.}~\bibnamefont {Leonhardt}},\ }\bibfield  {title} {\bibinfo {title} {Towards 10 {{meV}} resolution: {{The}} design of an ultrahigh resolution soft {{X-ray RIXS}} spectrometer},\ }\href {https://doi.org/10.1063/1.4964847} {\bibfield  {journal} {\bibinfo  {journal} {Review of Scientific Instruments}\ }\textbf {\bibinfo {volume} {87}},\ \bibinfo {pages} {115109} (\bibinfo {year} {2016})}\BibitemShut {NoStop}%
\bibitem [{Sup()}]{SuppMat}%
  \BibitemOpen
  \href@noop {} {}\bibinfo {note} {See Supplemental Material at [URL will be inserted by publisher] for more details of the experiments and calculations.}\BibitemShut {Stop}%
\bibitem [{\citenamefont {Scheie}(2021)}]{PyCrystalField}%
  \BibitemOpen
  \bibfield  {author} {\bibinfo {author} {\bibfnamefont {A.}~\bibnamefont {Scheie}},\ }\bibfield  {title} {\bibinfo {title} {Pycrystalfield: software for calculation, analysis and fitting of crystal electric field hamiltonians},\ }\href {https://doi.org/10.1107/S160057672001554X} {\bibfield  {journal} {\bibinfo  {journal} {Journal of Applied Crystallography}\ }\textbf {\bibinfo {volume} {54}} (\bibinfo {year} {2021})}\BibitemShut {NoStop}%
\bibitem [{\citenamefont {Stevens}(1952)}]{Stevens1952}%
  \BibitemOpen
  \bibfield  {author} {\bibinfo {author} {\bibfnamefont {K.~W.~H.}\ \bibnamefont {Stevens}},\ }\bibfield  {title} {\bibinfo {title} {Matrix elements and operator equivalents connected with the magnetic properties of rare earth ions},\ }\href {http://stacks.iop.org/0370-1298/65/i=3/a=308} {\bibfield  {journal} {\bibinfo  {journal} {Proceedings of the Physical Society. Section A}\ }\textbf {\bibinfo {volume} {65}},\ \bibinfo {pages} {209} (\bibinfo {year} {1952})}\BibitemShut {NoStop}%
\bibitem [{\citenamefont {Hansmann}\ \emph {et~al.}(2008)\citenamefont {Hansmann}, \citenamefont {Severing}, \citenamefont {Hu}, \citenamefont {Haverkort}, \citenamefont {Chang}, \citenamefont {Klein}, \citenamefont {Tanaka}, \citenamefont {Hsieh}, \citenamefont {Lin}, \citenamefont {Chen}, \citenamefont {F{\aa}k}, \citenamefont {Lejay},\ and\ \citenamefont {Tjeng}}]{2008_Hansmann_PhysRevLett}%
  \BibitemOpen
  \bibfield  {author} {\bibinfo {author} {\bibfnamefont {P.}~\bibnamefont {Hansmann}}, \bibinfo {author} {\bibfnamefont {A.}~\bibnamefont {Severing}}, \bibinfo {author} {\bibfnamefont {Z.}~\bibnamefont {Hu}}, \bibinfo {author} {\bibfnamefont {M.~W.}\ \bibnamefont {Haverkort}}, \bibinfo {author} {\bibfnamefont {C.~F.}\ \bibnamefont {Chang}}, \bibinfo {author} {\bibfnamefont {S.}~\bibnamefont {Klein}}, \bibinfo {author} {\bibfnamefont {A.}~\bibnamefont {Tanaka}}, \bibinfo {author} {\bibfnamefont {H.~H.}\ \bibnamefont {Hsieh}}, \bibinfo {author} {\bibfnamefont {H.-J.}\ \bibnamefont {Lin}}, \bibinfo {author} {\bibfnamefont {C.~T.}\ \bibnamefont {Chen}}, \bibinfo {author} {\bibfnamefont {B.}~\bibnamefont {F{\aa}k}}, \bibinfo {author} {\bibfnamefont {P.}~\bibnamefont {Lejay}},\ and\ \bibinfo {author} {\bibfnamefont {L.~H.}\ \bibnamefont {Tjeng}},\ }\bibfield  {title} {\bibinfo {title} {Determining the {{Crystal-Field Ground State}} in {{Rare Earth Heavy Fermion Materials Using Soft-X-Ray Absorption
  Spectroscopy}}},\ }\href {https://doi.org/10.1103/PhysRevLett.100.066405} {\bibfield  {journal} {\bibinfo  {journal} {Physical Review Letters}\ }\textbf {\bibinfo {volume} {100}},\ \bibinfo {pages} {066405} (\bibinfo {year} {2008})}\BibitemShut {NoStop}%
\bibitem [{\citenamefont {Willers}\ \emph {et~al.}(2015)\citenamefont {Willers}, \citenamefont {Strigari}, \citenamefont {Hu}, \citenamefont {Sessi}, \citenamefont {Brookes}, \citenamefont {Bauer}, \citenamefont {Sarrao}, \citenamefont {Thompson}, \citenamefont {Tanaka}, \citenamefont {Wirth}, \citenamefont {Tjeng},\ and\ \citenamefont {Severing}}]{2015_Willers_ProcNatlAcadSci}%
  \BibitemOpen
  \bibfield  {author} {\bibinfo {author} {\bibfnamefont {T.}~\bibnamefont {Willers}}, \bibinfo {author} {\bibfnamefont {F.}~\bibnamefont {Strigari}}, \bibinfo {author} {\bibfnamefont {Z.}~\bibnamefont {Hu}}, \bibinfo {author} {\bibfnamefont {V.}~\bibnamefont {Sessi}}, \bibinfo {author} {\bibfnamefont {N.~B.}\ \bibnamefont {Brookes}}, \bibinfo {author} {\bibfnamefont {E.~D.}\ \bibnamefont {Bauer}}, \bibinfo {author} {\bibfnamefont {J.~L.}\ \bibnamefont {Sarrao}}, \bibinfo {author} {\bibfnamefont {J.~D.}\ \bibnamefont {Thompson}}, \bibinfo {author} {\bibfnamefont {A.}~\bibnamefont {Tanaka}}, \bibinfo {author} {\bibfnamefont {S.}~\bibnamefont {Wirth}}, \bibinfo {author} {\bibfnamefont {L.~H.}\ \bibnamefont {Tjeng}},\ and\ \bibinfo {author} {\bibfnamefont {A.}~\bibnamefont {Severing}},\ }\bibfield  {title} {\bibinfo {title} {Correlation between ground state and orbital anisotropy in heavy fermion materials},\ }\href {https://doi.org/10.1073/pnas.1415657112} {\bibfield  {journal} {\bibinfo  {journal} {Proceedings
  of the National Academy of Sciences}\ }\textbf {\bibinfo {volume} {112}},\ \bibinfo {pages} {2384} (\bibinfo {year} {2015})}\BibitemShut {NoStop}%
\bibitem [{\citenamefont {Christovam}\ \emph {et~al.}(2024)\citenamefont {Christovam}, \citenamefont {{Ferreira-Carvalho}}, \citenamefont {Marino}, \citenamefont {Sundermann}, \citenamefont {Takegami}, \citenamefont {{Melendez-Sans}}, \citenamefont {Tsuei}, \citenamefont {Hu}, \citenamefont {R{\"o}{\ss}ler}, \citenamefont {Valvidares}, \citenamefont {Haverkort}, \citenamefont {Liu}, \citenamefont {Bauer}, \citenamefont {Tjeng}, \citenamefont {Zwicknagl},\ and\ \citenamefont {Severing}}]{2024_Christovam_PhysRevLett}%
  \BibitemOpen
  \bibfield  {author} {\bibinfo {author} {\bibfnamefont {D.~S.}\ \bibnamefont {Christovam}}, \bibinfo {author} {\bibfnamefont {M.}~\bibnamefont {{Ferreira-Carvalho}}}, \bibinfo {author} {\bibfnamefont {A.}~\bibnamefont {Marino}}, \bibinfo {author} {\bibfnamefont {M.}~\bibnamefont {Sundermann}}, \bibinfo {author} {\bibfnamefont {D.}~\bibnamefont {Takegami}}, \bibinfo {author} {\bibfnamefont {A.}~\bibnamefont {{Melendez-Sans}}}, \bibinfo {author} {\bibfnamefont {K.~D.}\ \bibnamefont {Tsuei}}, \bibinfo {author} {\bibfnamefont {Z.}~\bibnamefont {Hu}}, \bibinfo {author} {\bibfnamefont {S.}~\bibnamefont {R{\"o}{\ss}ler}}, \bibinfo {author} {\bibfnamefont {M.}~\bibnamefont {Valvidares}}, \bibinfo {author} {\bibfnamefont {M.~W.}\ \bibnamefont {Haverkort}}, \bibinfo {author} {\bibfnamefont {Y.}~\bibnamefont {Liu}}, \bibinfo {author} {\bibfnamefont {E.~D.}\ \bibnamefont {Bauer}}, \bibinfo {author} {\bibfnamefont {L.~H.}\ \bibnamefont {Tjeng}}, \bibinfo {author} {\bibfnamefont {G.}~\bibnamefont {Zwicknagl}},\ and\
  \bibinfo {author} {\bibfnamefont {A.}~\bibnamefont {Severing}},\ }\bibfield  {title} {\bibinfo {title} {Spectroscopic {{Evidence}} of {{Kondo-Induced Quasiquartet}} in {{Ce}}{{Rh}}$_2${{As}}$_2$},\ }\href {https://doi.org/10.1103/PhysRevLett.132.046401} {\bibfield  {journal} {\bibinfo  {journal} {Physical Review Letters}\ }\textbf {\bibinfo {volume} {132}},\ \bibinfo {pages} {046401} (\bibinfo {year} {2024})}\BibitemShut {NoStop}%
\bibitem [{\citenamefont {Haverkort}\ \emph {et~al.}(2012)\citenamefont {Haverkort}, \citenamefont {Zwierzycki},\ and\ \citenamefont {Andersen}}]{2012_Haverkort_PhysRevB}%
  \BibitemOpen
  \bibfield  {author} {\bibinfo {author} {\bibfnamefont {M.~W.}\ \bibnamefont {Haverkort}}, \bibinfo {author} {\bibfnamefont {M.}~\bibnamefont {Zwierzycki}},\ and\ \bibinfo {author} {\bibfnamefont {O.~K.}\ \bibnamefont {Andersen}},\ }\bibfield  {title} {\bibinfo {title} {Multiplet ligand-field theory using {{Wannier}} orbitals},\ }\href {https://doi.org/10.1103/PhysRevB.85.165113} {\bibfield  {journal} {\bibinfo  {journal} {Physical Review B}\ }\textbf {\bibinfo {volume} {85}},\ \bibinfo {pages} {165113} (\bibinfo {year} {2012})}\BibitemShut {NoStop}%
\bibitem [{\citenamefont {Amorese}\ \emph {et~al.}(2016)\citenamefont {Amorese}, \citenamefont {Dellea}, \citenamefont {Fanciulli}, \citenamefont {Seiro}, \citenamefont {Geibel}, \citenamefont {Krellner}, \citenamefont {Makarova}, \citenamefont {Braicovich}, \citenamefont {Ghiringhelli}, \citenamefont {Vyalikh}, \citenamefont {Brookes},\ and\ \citenamefont {Kummer}}]{2016_Amorese_PhysRevB}%
  \BibitemOpen
  \bibfield  {author} {\bibinfo {author} {\bibfnamefont {A.}~\bibnamefont {Amorese}}, \bibinfo {author} {\bibfnamefont {G.}~\bibnamefont {Dellea}}, \bibinfo {author} {\bibfnamefont {M.}~\bibnamefont {Fanciulli}}, \bibinfo {author} {\bibfnamefont {S.}~\bibnamefont {Seiro}}, \bibinfo {author} {\bibfnamefont {C.}~\bibnamefont {Geibel}}, \bibinfo {author} {\bibfnamefont {C.}~\bibnamefont {Krellner}}, \bibinfo {author} {\bibfnamefont {I.~P.}\ \bibnamefont {Makarova}}, \bibinfo {author} {\bibfnamefont {L.}~\bibnamefont {Braicovich}}, \bibinfo {author} {\bibfnamefont {G.}~\bibnamefont {Ghiringhelli}}, \bibinfo {author} {\bibfnamefont {D.~V.}\ \bibnamefont {Vyalikh}}, \bibinfo {author} {\bibfnamefont {N.~B.}\ \bibnamefont {Brookes}},\ and\ \bibinfo {author} {\bibfnamefont {K.}~\bibnamefont {Kummer}},\ }\bibfield  {title} {\bibinfo {title} {4$f$ excitations in {{Ce Kondo}} lattices studied by resonant inelastic x-ray scattering},\ }\href {https://doi.org/10.1103/PhysRevB.93.165134} {\bibfield  {journal} {\bibinfo
  {journal} {Physical Review B}\ }\textbf {\bibinfo {volume} {93}},\ \bibinfo {pages} {165134} (\bibinfo {year} {2016})}\BibitemShut {NoStop}%
\bibitem [{\citenamefont {Horn}\ \emph {et~al.}(1981)\citenamefont {Horn}, \citenamefont {{Holland-Moritz}}, \citenamefont {Loewenhaupt}, \citenamefont {Steglich}, \citenamefont {Scheuer}, \citenamefont {Benoit},\ and\ \citenamefont {Flouquet}}]{1981_Horn_PhysRevB}%
  \BibitemOpen
  \bibfield  {author} {\bibinfo {author} {\bibfnamefont {S.}~\bibnamefont {Horn}}, \bibinfo {author} {\bibfnamefont {E.}~\bibnamefont {{Holland-Moritz}}}, \bibinfo {author} {\bibfnamefont {M.}~\bibnamefont {Loewenhaupt}}, \bibinfo {author} {\bibfnamefont {F.}~\bibnamefont {Steglich}}, \bibinfo {author} {\bibfnamefont {H.}~\bibnamefont {Scheuer}}, \bibinfo {author} {\bibfnamefont {A.}~\bibnamefont {Benoit}},\ and\ \bibinfo {author} {\bibfnamefont {J.}~\bibnamefont {Flouquet}},\ }\bibfield  {title} {\bibinfo {title} {Magnetic neutron scattering and crystal-field states in {Ce}{Cu}$_2${Si}$_2$},\ }\href {https://doi.org/10.1103/PhysRevB.23.3171} {\bibfield  {journal} {\bibinfo  {journal} {Physical Review B}\ }\textbf {\bibinfo {volume} {23}},\ \bibinfo {pages} {3171} (\bibinfo {year} {1981})}\BibitemShut {NoStop}%
\bibitem [{\citenamefont {Erkelens}\ and\ \citenamefont {Leiden~(Netherlands)}(1987)}]{1987_Erkelens_}%
  \BibitemOpen
  \bibfield  {author} {\bibinfo {author} {\bibfnamefont {W.~a.~C.}\ \bibnamefont {Erkelens}}\ and\ \bibinfo {author} {\bibfnamefont {R.}~\bibnamefont {Leiden~(Netherlands)}},\ }\href@noop {} {\emph {\bibinfo {title} {Elastic and Inelastic Neutron Scattering Studies on 3d and 4f Magnetic Compounds}}},\ \bibinfo {type} {Tech. Rep.}\ \bibinfo {number} {INIS-mf--11365}\ (\bibinfo  {institution} {Rijksuniversiteit Leiden (Netherlands)},\ \bibinfo {year} {1987})\BibitemShut {NoStop}%
\bibitem [{\citenamefont {Grier}\ \emph {et~al.}(1988)\citenamefont {Grier}, \citenamefont {Lawrence}, \citenamefont {Horn},\ and\ \citenamefont {Thompson}}]{1988_Grier_JPhysCSolidStatePhys}%
  \BibitemOpen
  \bibfield  {author} {\bibinfo {author} {\bibfnamefont {B.~H.}\ \bibnamefont {Grier}}, \bibinfo {author} {\bibfnamefont {J.~M.}\ \bibnamefont {Lawrence}}, \bibinfo {author} {\bibfnamefont {S.}~\bibnamefont {Horn}},\ and\ \bibinfo {author} {\bibfnamefont {J.~D.}\ \bibnamefont {Thompson}},\ }\bibfield  {title} {\bibinfo {title} {Inelastic magnetic neutron scattering in {{CeT$_2$Si$_2$}} ({{T}}={{Au}}, {{Pd}}, {{Ru}} or {{Pt}}): Crystal fields and quasi-elastic scattering},\ }\href {https://doi.org/10.1088/0022-3719/21/6/016} {\bibfield  {journal} {\bibinfo  {journal} {Journal of Physics C: Solid State Physics}\ }\textbf {\bibinfo {volume} {21}},\ \bibinfo {pages} {1099} (\bibinfo {year} {1988})}\BibitemShut {NoStop}%
\bibitem [{\citenamefont {Litvin}(1974)}]{1974_Litvin_Physica}%
  \BibitemOpen
  \bibfield  {author} {\bibinfo {author} {\bibfnamefont {D.~B.}\ \bibnamefont {Litvin}},\ }\bibfield  {title} {\bibinfo {title} {The {{Luttinger-Tisza}} method},\ }\href {https://doi.org/10.1016/0031-8914(74)90257-2} {\bibfield  {journal} {\bibinfo  {journal} {Physica}\ }\textbf {\bibinfo {volume} {77}},\ \bibinfo {pages} {205} (\bibinfo {year} {1974})}\BibitemShut {NoStop}%
\bibitem [{\citenamefont {Vijayvargia}\ \emph {et~al.}(2024)\citenamefont {Vijayvargia}, \citenamefont {Zhang}, \citenamefont {Barros}, \citenamefont {Lin},\ and\ \citenamefont {Erten}}]{2024_Vijayvargia_}%
  \BibitemOpen
  \bibfield  {author} {\bibinfo {author} {\bibfnamefont {A.}~\bibnamefont {Vijayvargia}}, \bibinfo {author} {\bibfnamefont {H.}~\bibnamefont {Zhang}}, \bibinfo {author} {\bibfnamefont {K.}~\bibnamefont {Barros}}, \bibinfo {author} {\bibfnamefont {S.-Z.}\ \bibnamefont {Lin}},\ and\ \bibinfo {author} {\bibfnamefont {O.}~\bibnamefont {Erten}},\ }\href {https://doi.org/10.48550/arXiv.2410.06344} {\bibinfo {title} {Electric field driven spin textures in heavy fermion van der {{Waals}} magnets}} (\bibinfo {year} {2024}),\ \Eprint {https://arxiv.org/abs/2410.06344} {arXiv:2410.06344 [cond-mat]} \BibitemShut {NoStop}%
\bibitem [{\citenamefont {Dahlbom}\ \emph {et~al.}(2025)\citenamefont {Dahlbom}, \citenamefont {Zhang}, \citenamefont {Miles}, \citenamefont {Quinn}, \citenamefont {Niraula}, \citenamefont {Thipe}, \citenamefont {Wilson}, \citenamefont {Matin}, \citenamefont {Mankad}, \citenamefont {Hahn}, \citenamefont {Pajerowski}, \citenamefont {Johnston}, \citenamefont {Wang}, \citenamefont {Lane}, \citenamefont {Li}, \citenamefont {Bai}, \citenamefont {Mourigal}, \citenamefont {Batista},\ and\ \citenamefont {Barros}}]{2025_Dahlbom_}%
  \BibitemOpen
  \bibfield  {author} {\bibinfo {author} {\bibfnamefont {D.}~\bibnamefont {Dahlbom}}, \bibinfo {author} {\bibfnamefont {H.}~\bibnamefont {Zhang}}, \bibinfo {author} {\bibfnamefont {C.}~\bibnamefont {Miles}}, \bibinfo {author} {\bibfnamefont {S.}~\bibnamefont {Quinn}}, \bibinfo {author} {\bibfnamefont {A.}~\bibnamefont {Niraula}}, \bibinfo {author} {\bibfnamefont {B.}~\bibnamefont {Thipe}}, \bibinfo {author} {\bibfnamefont {M.}~\bibnamefont {Wilson}}, \bibinfo {author} {\bibfnamefont {S.}~\bibnamefont {Matin}}, \bibinfo {author} {\bibfnamefont {H.}~\bibnamefont {Mankad}}, \bibinfo {author} {\bibfnamefont {S.}~\bibnamefont {Hahn}}, \bibinfo {author} {\bibfnamefont {D.}~\bibnamefont {Pajerowski}}, \bibinfo {author} {\bibfnamefont {S.}~\bibnamefont {Johnston}}, \bibinfo {author} {\bibfnamefont {Z.}~\bibnamefont {Wang}}, \bibinfo {author} {\bibfnamefont {H.}~\bibnamefont {Lane}}, \bibinfo {author} {\bibfnamefont {Y.~W.}\ \bibnamefont {Li}}, \bibinfo {author} {\bibfnamefont {X.}~\bibnamefont {Bai}}, \bibinfo
  {author} {\bibfnamefont {M.}~\bibnamefont {Mourigal}}, \bibinfo {author} {\bibfnamefont {C.~D.}\ \bibnamefont {Batista}},\ and\ \bibinfo {author} {\bibfnamefont {K.}~\bibnamefont {Barros}},\ }\href {https://doi.org/10.48550/arXiv.2501.13095} {\bibinfo {title} {Sunny.jl: {{A Julia Package}} for {{Spin Dynamics}}}} (\bibinfo {year} {2025}),\ \Eprint {https://arxiv.org/abs/2501.13095} {arXiv:2501.13095 [quant-ph]} \BibitemShut {NoStop}%
\bibitem [{\citenamefont {Allen}\ and\ \citenamefont {Young}(1977)}]{1977_Allen_PhysRevB}%
  \BibitemOpen
  \bibfield  {author} {\bibinfo {author} {\bibfnamefont {J.~W.}\ \bibnamefont {Allen}}\ and\ \bibinfo {author} {\bibfnamefont {C.~Y.}\ \bibnamefont {Young}},\ }\bibfield  {title} {\bibinfo {title} {Magnetic anisotropy due to spin-orbit and dipole-dipole interactions in spin-density-wave antiferromagnets},\ }\href {https://doi.org/10.1103/PhysRevB.16.1103} {\bibfield  {journal} {\bibinfo  {journal} {Physical Review B}\ }\textbf {\bibinfo {volume} {16}},\ \bibinfo {pages} {1103} (\bibinfo {year} {1977})}\BibitemShut {NoStop}%
\bibitem [{\citenamefont {Chen}\ \emph {et~al.}(2013)\citenamefont {Chen}, \citenamefont {Ju}, \citenamefont {Jiang}, \citenamefont {Starykh},\ and\ \citenamefont {Balents}}]{2013_Chen_PhysRevB}%
  \BibitemOpen
  \bibfield  {author} {\bibinfo {author} {\bibfnamefont {R.}~\bibnamefont {Chen}}, \bibinfo {author} {\bibfnamefont {H.}~\bibnamefont {Ju}}, \bibinfo {author} {\bibfnamefont {H.-C.}\ \bibnamefont {Jiang}}, \bibinfo {author} {\bibfnamefont {O.~A.}\ \bibnamefont {Starykh}},\ and\ \bibinfo {author} {\bibfnamefont {L.}~\bibnamefont {Balents}},\ }\bibfield  {title} {\bibinfo {title} {Ground states of spin-$\frac{1}{2}$ triangular antiferromagnets in a magnetic field},\ }\href {https://doi.org/10.1103/PhysRevB.87.165123} {\bibfield  {journal} {\bibinfo  {journal} {Physical Review B}\ }\textbf {\bibinfo {volume} {87}},\ \bibinfo {pages} {165123} (\bibinfo {year} {2013})}\BibitemShut {NoStop}%
\bibitem [{\citenamefont {Facheris}\ \emph {et~al.}(2022)\citenamefont {Facheris}, \citenamefont {Povarov}, \citenamefont {Nabi}, \citenamefont {Mazzone}, \citenamefont {Lass}, \citenamefont {Roessli}, \citenamefont {Ressouche}, \citenamefont {Yan}, \citenamefont {Gvasaliya},\ and\ \citenamefont {Zheludev}}]{2022_Facheris_PhysRevLett}%
  \BibitemOpen
  \bibfield  {author} {\bibinfo {author} {\bibfnamefont {L.}~\bibnamefont {Facheris}}, \bibinfo {author} {\bibfnamefont {K.~{\relax Yu}.}\ \bibnamefont {Povarov}}, \bibinfo {author} {\bibfnamefont {S.~D.}\ \bibnamefont {Nabi}}, \bibinfo {author} {\bibfnamefont {D.~G.}\ \bibnamefont {Mazzone}}, \bibinfo {author} {\bibfnamefont {J.}~\bibnamefont {Lass}}, \bibinfo {author} {\bibfnamefont {B.}~\bibnamefont {Roessli}}, \bibinfo {author} {\bibfnamefont {E.}~\bibnamefont {Ressouche}}, \bibinfo {author} {\bibfnamefont {Z.}~\bibnamefont {Yan}}, \bibinfo {author} {\bibfnamefont {S.}~\bibnamefont {Gvasaliya}},\ and\ \bibinfo {author} {\bibfnamefont {A.}~\bibnamefont {Zheludev}},\ }\bibfield  {title} {\bibinfo {title} {Spin {{Density Wave}} versus {{Fractional Magnetization Plateau}} in a {{Triangular Antiferromagnet}}},\ }\href {https://doi.org/10.1103/PhysRevLett.129.087201} {\bibfield  {journal} {\bibinfo  {journal} {Physical Review Letters}\ }\textbf {\bibinfo {volume} {129}},\ \bibinfo {pages} {087201} (\bibinfo
  {year} {2022})}\BibitemShut {NoStop}%
\bibitem [{\citenamefont {Park}\ \emph {et~al.}(2025)\citenamefont {Park}, \citenamefont {Ortiz}, \citenamefont {Sprague}, \citenamefont {Sakhya}, \citenamefont {Chen}, \citenamefont {Frontzek}, \citenamefont {Tian}, \citenamefont {Sibille}, \citenamefont {Mazzone}, \citenamefont {Tabata}, \citenamefont {Kaneko}, \citenamefont {{DeBeer-Schmitt}}, \citenamefont {Stone}, \citenamefont {Parker}, \citenamefont {Samolyuk}, \citenamefont {Miao}, \citenamefont {Neupane},\ and\ \citenamefont {Christianson}}]{2025_Park_NatCommun}%
  \BibitemOpen
  \bibfield  {author} {\bibinfo {author} {\bibfnamefont {P.}~\bibnamefont {Park}}, \bibinfo {author} {\bibfnamefont {B.~R.}\ \bibnamefont {Ortiz}}, \bibinfo {author} {\bibfnamefont {M.}~\bibnamefont {Sprague}}, \bibinfo {author} {\bibfnamefont {A.~P.}\ \bibnamefont {Sakhya}}, \bibinfo {author} {\bibfnamefont {S.~A.}\ \bibnamefont {Chen}}, \bibinfo {author} {\bibfnamefont {M.~D.}\ \bibnamefont {Frontzek}}, \bibinfo {author} {\bibfnamefont {W.}~\bibnamefont {Tian}}, \bibinfo {author} {\bibfnamefont {R.}~\bibnamefont {Sibille}}, \bibinfo {author} {\bibfnamefont {D.~G.}\ \bibnamefont {Mazzone}}, \bibinfo {author} {\bibfnamefont {C.}~\bibnamefont {Tabata}}, \bibinfo {author} {\bibfnamefont {K.}~\bibnamefont {Kaneko}}, \bibinfo {author} {\bibfnamefont {L.~M.}\ \bibnamefont {{DeBeer-Schmitt}}}, \bibinfo {author} {\bibfnamefont {M.~B.}\ \bibnamefont {Stone}}, \bibinfo {author} {\bibfnamefont {D.~S.}\ \bibnamefont {Parker}}, \bibinfo {author} {\bibfnamefont {G.~D.}\ \bibnamefont {Samolyuk}}, \bibinfo {author}
  {\bibfnamefont {H.}~\bibnamefont {Miao}}, \bibinfo {author} {\bibfnamefont {M.}~\bibnamefont {Neupane}},\ and\ \bibinfo {author} {\bibfnamefont {A.~D.}\ \bibnamefont {Christianson}},\ }\bibfield  {title} {\bibinfo {title} {Spin density wave and van {{Hove}} singularity in the kagome metal {{CeTi$_3$Bi$_4$}}},\ }\href {https://doi.org/10.1038/s41467-025-59460-4} {\bibfield  {journal} {\bibinfo  {journal} {Nature Communications}\ }\textbf {\bibinfo {volume} {16}},\ \bibinfo {pages} {4384} (\bibinfo {year} {2025})}\BibitemShut {NoStop}%
\bibitem [{\citenamefont {Starykh}\ \emph {et~al.}(2010)\citenamefont {Starykh}, \citenamefont {Katsura},\ and\ \citenamefont {Balents}}]{2010_Starykh_PhysRevB}%
  \BibitemOpen
  \bibfield  {author} {\bibinfo {author} {\bibfnamefont {O.~A.}\ \bibnamefont {Starykh}}, \bibinfo {author} {\bibfnamefont {H.}~\bibnamefont {Katsura}},\ and\ \bibinfo {author} {\bibfnamefont {L.}~\bibnamefont {Balents}},\ }\bibfield  {title} {\bibinfo {title} {Extreme sensitivity of a frustrated quantum magnet: {{Cs}}$_2${{Cu}}{{Cl}}$_4$},\ }\href {https://doi.org/10.1103/PhysRevB.82.014421} {\bibfield  {journal} {\bibinfo  {journal} {Physical Review B}\ }\textbf {\bibinfo {volume} {82}},\ \bibinfo {pages} {014421} (\bibinfo {year} {2010})}\BibitemShut {NoStop}%
\bibitem [{\citenamefont {Dahlbom}(2022)}]{2022_Dahlbom_PhysRevB}%
  \BibitemOpen
  \bibfield  {author} {\bibinfo {author} {\bibfnamefont {D.}~\bibnamefont {Dahlbom}},\ }\bibfield  {title} {\bibinfo {title} {Langevin dynamics of generalized spins as coherent states {{S}}{{U}}({{$N$}})},\ }\bibfield  {journal} {\bibinfo  {journal} {Physical Review B}\ }\textbf {\bibinfo {volume} {106}},\ \href {https://doi.org/10.1103/PhysRevB.106.235154} {10.1103/PhysRevB.106.235154} (\bibinfo {year} {2022})\BibitemShut {NoStop}%
\bibitem [{\citenamefont {Chubukov}\ and\ \citenamefont {Golosov}(1991)}]{Chubukov_1991}%
  \BibitemOpen
  \bibfield  {author} {\bibinfo {author} {\bibfnamefont {A.~V.}\ \bibnamefont {Chubukov}}\ and\ \bibinfo {author} {\bibfnamefont {D.~I.}\ \bibnamefont {Golosov}},\ }\bibfield  {title} {\bibinfo {title} {Quantum theory of an antiferromagnet on a triangular lattice in a magnetic field},\ }\href {https://doi.org/10.1088/0953-8984/3/1/005} {\bibfield  {journal} {\bibinfo  {journal} {Journal of Physics: Condensed Matter}\ }\textbf {\bibinfo {volume} {3}},\ \bibinfo {pages} {69} (\bibinfo {year} {1991})}\BibitemShut {NoStop}%
\bibitem [{\citenamefont {Coletta}\ \emph {et~al.}(2016)\citenamefont {Coletta}, \citenamefont {T\'oth}, \citenamefont {Penc},\ and\ \citenamefont {Mila}}]{PhysRevB.94.075136}%
  \BibitemOpen
  \bibfield  {author} {\bibinfo {author} {\bibfnamefont {T.}~\bibnamefont {Coletta}}, \bibinfo {author} {\bibfnamefont {T.~A.}\ \bibnamefont {T\'oth}}, \bibinfo {author} {\bibfnamefont {K.}~\bibnamefont {Penc}},\ and\ \bibinfo {author} {\bibfnamefont {F.}~\bibnamefont {Mila}},\ }\bibfield  {title} {\bibinfo {title} {Semiclassical theory of the magnetization process of the triangular lattice {{Heisenberg}} model},\ }\href {https://doi.org/10.1103/PhysRevB.94.075136} {\bibfield  {journal} {\bibinfo  {journal} {Phys. Rev. B}\ }\textbf {\bibinfo {volume} {94}},\ \bibinfo {pages} {075136} (\bibinfo {year} {2016})}\BibitemShut {NoStop}%
\end{thebibliography}
\end{document}


\sloppy
\title{Supplemental material for: Magnetic order, magnons, and crystal fields in van-der-Waals CeSiI}

\author{Wolfgang Simeth$^1$}
\author{Connor A. Occhialini$^{2,3}$}
\author{Michael E. Ziebel$^{2,4}$}
\author{Nethmi W. Hewage$^4$}
\author{Sabrina J. Li$^{5,6}$}
\author{Daniel Pajerowski$^7$}
\author{Taehun Kim$^{2,3}$}
\author{Ben Zager$^{3,8}$}
\author{Jonathan Pelliciari$^{3}$}
\author{Kipton Barros$^9$}
\author{Daniel Rehn$^{5}$} 
\author{Abhay N. Pasupathy$^{2,8}$}
\author{Valentina Bisogni$^3$}
\author{Xavier Roy$^4$}
\author{Allen Scheie$^1$}

\affiliation{$^1$MPA-Q, Los Alamos National Laboratory, Los Alamos, New Mexico 87545, USA}
\affiliation{$^2$Department of Physics, Columbia University, New York, New York 10027, USA}
\affiliation{$^3$National Synchrotron Light Source II, Brookhaven National Laboratory, Upton, NY 11973, USA}
\affiliation{$^4$Department of Chemistry, Columbia University, New York, NY, USA}
\affiliation{$^5$Computational Physics Division, Los Alamos National Laboratory, Los Alamos, New Mexico 87545}
\affiliation{$^6$Center for Nonlinear Studies, Los Alamos National Laboratory, Los Alamos, New Mexico 87545}
\affiliation{$^7$Neutron Scattering Division, Oak Ridge National Laboratory, Oak Ridge, TN 37831, USA}
\affiliation{$^8$Department of Condensed Matter Physics and Materials Science, Brookhaven National Laboratory, Upton, New York 11973, USA}
\affiliation{$^9$Theoretical Division and CNLS, Los Alamos National Laboratory, Los Alamos, New Mexico 87545, USA}

\date{\today}


\maketitle

\newpage

\section{Neutron spectroscopy}
\begin{figure}[b!]
    \includegraphics[]{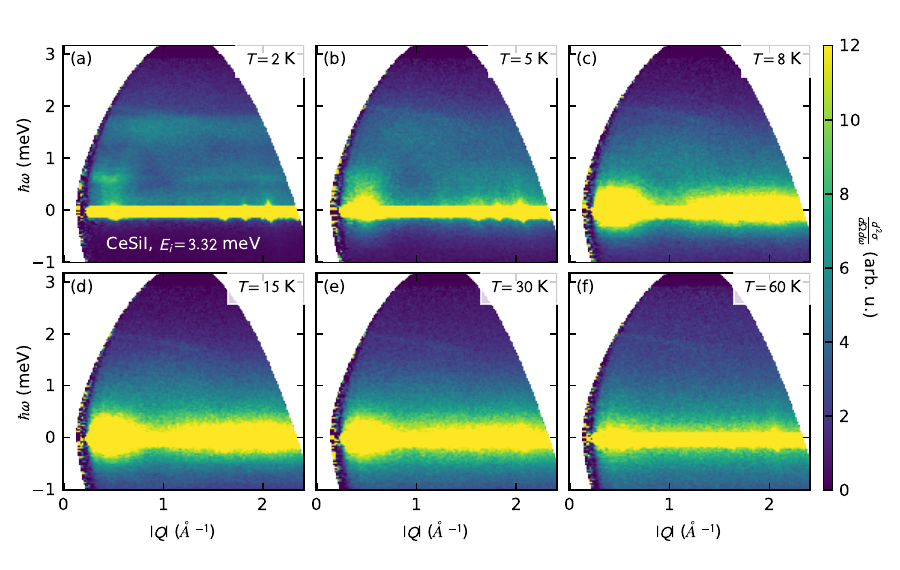} 
    \caption{Thermal variation of neutron powder spectroscopy, taken with incident neutron energy $E_i=3.32\,$meV. Panels (a)-(f) show data taken at different temperatures.}  
    \label{fig:FigureSI:TemperatureDependence3meV}
\end{figure}

\begin{figure}
    \includegraphics[]{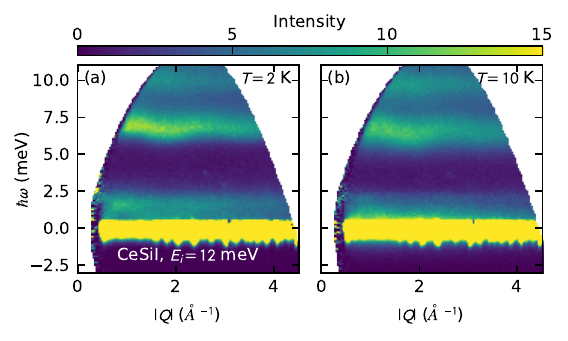} 
    \caption{Neutron powder spectroscopy data, taken with incident neutron energy $E_i=12\,$meV at temperatures (a) $T=2\,$K and (b) $T=10\,$K.}  
    \label{fig:FigureSI:12meVDataTwoTemperatures}
\end{figure}

\begin{figure}
    \includegraphics[]{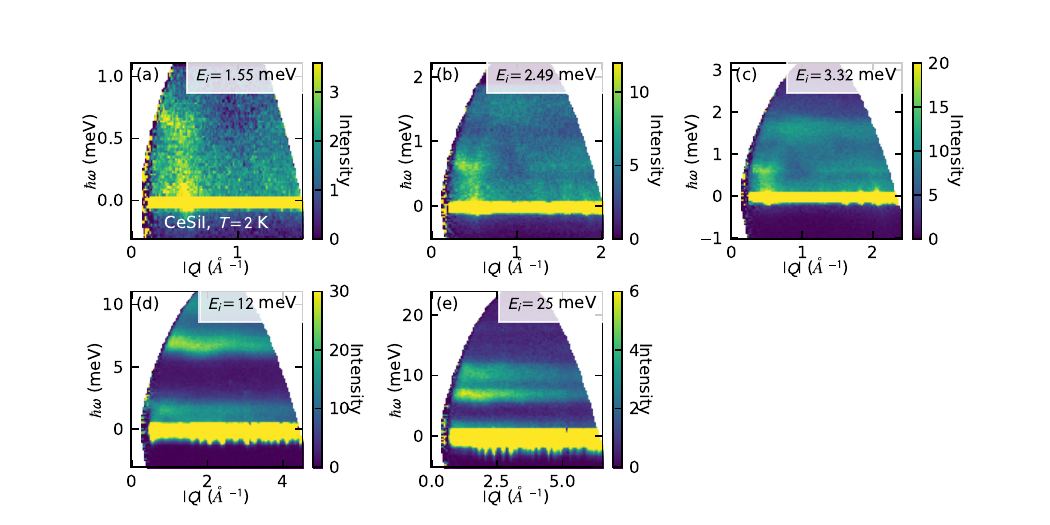} 
    \caption{Neutron powder spectroscopy data taken at $T=2\,$K with different incident energies.}  
    \label{fig:FigureSI:2KAllIncidentEnergies}
\end{figure}

We measured neutron spectroscopy using the CNCS spectrometer \cite{CNCS} at Oak Ridge National Laboratory's Spallation Neutron Source \cite{mason2006spallation}. We loaded 1.9~g CeSiI sample material in a Vanadium can. All data shown are subtracted by background, which was measured with an empty Vanadium can. 
We measured with incident neutron energies $E_i=1.55$~meV, 2.49~meV, 3.32~meV, 12~meV, and 25~meV.
Most measurements were performed at base temperature $T=2$~K, but we also measured the temperature dependence with $E_i=3.32$~meV at $T=2$~K, 5~K, 8~K, 15~K, 30~K, and 60~K. Data are presented in Fig.~\ref{fig:FigureSI:TemperatureDependence3meV}.
We also measured $E_i=25$~meV at both $T=2$~K and $T=60$~K [shown in main text Fig.~2(a) and (b), respectively] and $E_i=12$~meV at $T=2$~K and $T=10$~K
(shown in Fig.~\ref{fig:FigureSI:12meVDataTwoTemperatures}).
Figure~\ref{fig:FigureSI:2KAllIncidentEnergies} summarizes all neutron data taken at $T=2\,$K with incident energies (a) 1.55~meV, (b) 2.49~meV, (c) 12~meV, and (d) 25~meV. 

\clearpage
\newpage

\section{Crystal electric fields fits to neutron spectroscopy}

In the $E_i = 25$~meV neutron data shown in Fig.~2 of the manuscript, two crystal electric field (CEF) modes are visible: a well-defined mode at 6.8~meV, and a broad mode centered at 10.3~meV. 
These modes follow the same $Q$-dependence as the Ce$^{3+}$ magnetic form factor \cite{BrownFF}, and thus they clearly come from the Ce magnetism  [see Fig.~2(d) of main text]. 
Furthermore, these modes lie very close to the CEF level energies estimated from the specific heat ($7.5 \pm 0.5$~meV for the first excited level) \cite{posey2024CeSiI}, confirming they are the two CEF modes from magnetic Ce.

We used the CEF modes energies and relative intensities to fit a CEF model to the data. 
The crystal field Hamiltonian is written
\begin{equation}
\mathcal{H}_{CEF} = \sum_{n,m} B_n^m O_n^m 
\label{eq:CEF_H}
\end{equation}
where $B_n^m$ are scalar CEF parameters and $O_n^m$ are Stevens operators \cite{Stevens1952}. 
Because the Ce$^{3+}$ sites in CeSiI have three-fold rotation symmetry about $c$ and an effective $J=5/2$, if we set the quantization axis along $c$, symmetry dictates there are only three nonzero CEF parameters: $B_2^0$, $B_4^0$, and $B_4^3$.  
Using \texttt{PyCrystalField} \cite{PyCrystalField} software, we fit the $E_i = 25$~meV at $T=2$~K powder inelastic neutron data over $1\>{\text{\AA}^{-1}}< |Q| < 2\>{\text{\AA}^{-1}}$. 

\begin{figure*}
    \includegraphics[width=0.9\textwidth]{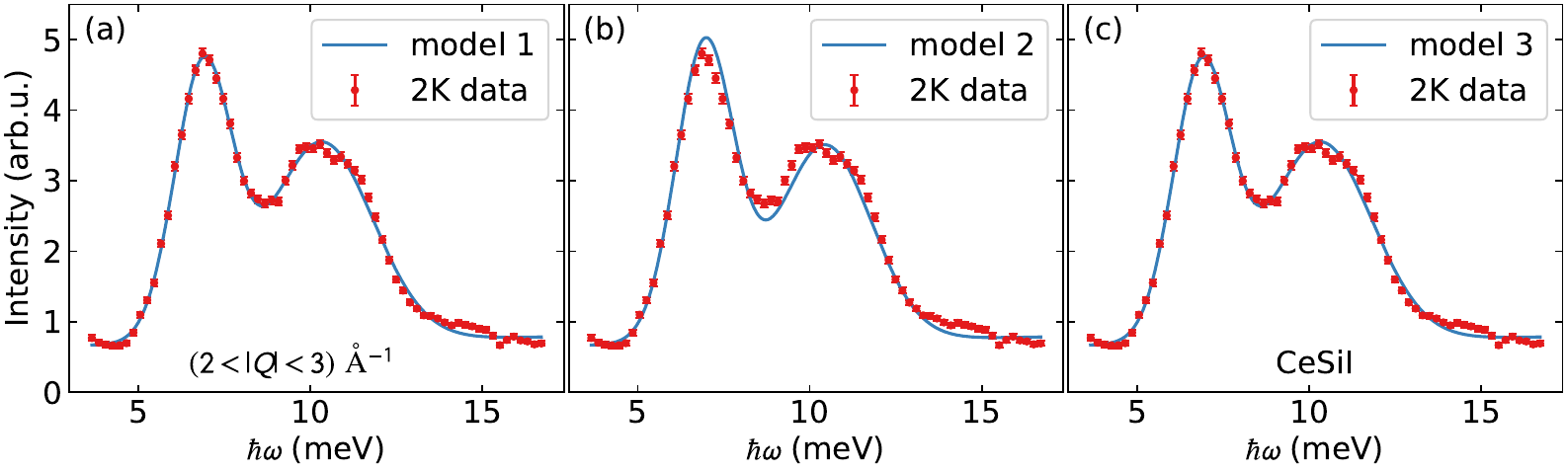} 
    \caption{Computed $T=2$~K spectra of the three CEF models in Table \ref{tab:CEFparams} compared with experimental data. Although model 2 in panel (b) visually fits the data the worst, XAS identifies it as the correct CEF model. Likely this discrepancy arises from the artifacts of phonon coupling which can modulate the intensities and lineshapes of CEF parameters \cite{Scheie_CEFphonon_2025}.}
    \label{fig:CEF-modelcomparison}
\end{figure*}

We began by calculating a point charge model \cite{Hutchings1964} assuming effective charges Si to be $-4e$ and I to be $+e$ (ionic values assuming charge balance in CeSiI), and then scaling the CEF parameters by 0.1 so the CEF levels lie near the predicted energies $\sim 10$~meV. These parameters are listed in Table \ref{tab:CEFparams}. 
Using these parameters as starting values, we fitted the three CEF parameters directly to the data in Fig. ~2(c) of the manuscript, including also an overall scale factor, to a $\chi^2_{red}$ loss function
\begin{equation}
    \chi^2_{red} = \frac{1}{\nu} \sum_i^N \frac{(y_i - f(\omega_i))^2}{\sigma_i^2}
\end{equation}
where $y_i$ are the experimental intensities at energy $\omega_i$ with statistical uncertainty $\sigma_i$, $f(\omega)$ is the theoretical CEF function, and $\nu = N - x$ is the residual degrees of freedom: $N$ data points minus $x$ fitted parameters. 
The calculated spectra from the best fit parameters of the three models are shown in Fig. \ref{fig:CEF-modelcomparison}.

Iterating the fit with many different starting values (beyond merely the point-charge estimate), we found three different solutions that fit the data almost equally well, listed in Table \ref{tab:CEFparams}. The Eigenvalues and eigenstates for each model are listed in Tables \ref{tab:Eigenvectors1}, \ref{tab:Eigenvectors2}, and \ref{tab:Eigenvectors3}. Uncertainty is determined using the method in Ref. \cite{Scheie_2022_CEFuncertainty}. (Interestingly, these three best fit models have a full permutation of the order of eigenstates: model 1 has the $J_z=\pm\frac{5}{2},\pm\frac{1}{2}$ doublets as the two lowest states and the $J_z=\pm\frac{3}{2}$ doublet as the highest state, model 2 has $J_z=\pm\frac{3}{2}$ as the middle eigenstate, and model 3 has $J_z=\pm\frac{3}{2}$ as the lowest eigenstate.)

\begin{table}[h]
\caption{CEF parameters of CeSiI from a point charge (PC) model and the three fitted models, along with the reduced $\chi^2$ from neutron scattering fits. Eigenspectra for the three models are in Tables \ref{tab:Eigenvectors1}, \ref{tab:Eigenvectors2}, and \ref{tab:Eigenvectors3}.}
\begin{ruledtabular}
\begin{tabular}{c|cccc}
CEF param. & PC (meV) & model 1 (meV) & model 2 (meV) & model 3 (meV)  \\ \hline
$B_{2}^{0}$  & $-0.622$  &  $ -0.306 \pm 0.016$  & $ -0.09 \pm 0.08$  &  $ -0.01 \pm 0.03$ \\
$B_{4}^{0}$  & $-0.0144$ &  $ -0.0221 \pm 0.0003$ &  $ -0.0055 \pm 0.0009$ &  $ 0.0319 \pm 0.0003$ \\
$B_{4}^{3}$  & $0.0061$ &  $ 0.288 \pm 0.014$  &  $  0.546 \pm 0.011$  &  $ 0.15 \pm 0.02$ \\ \hline
$\chi^2_{red}$ & & 3.31 & 7.60  & 3.31
\label{tab:CEFparams}
\end{tabular}\end{ruledtabular}
\label{tab:CEFparams}
\end{table}

\begin{table}[h]
\caption{Eigenvectors and Eigenvalues of the fitted CeSiI Hamiltonian for fitted model 1 (Table \ref{tab:CEFparams}).}
\begin{ruledtabular}
\begin{tabular}{c|cccccc}
E (meV) &$| -\frac{5}{2}\rangle$ & $| -\frac{3}{2}\rangle$ & $| -\frac{1}{2}\rangle$ & $| \frac{1}{2}\rangle$ & $| \frac{3}{2}\rangle$ & $| \frac{5}{2}\rangle$ \tabularnewline
 \hline 
0.0 & 0.0 & 0.0 & -0.44(3) & 0.0 & 0.0 & 0.897(13) \tabularnewline
0.0 & -0.897(13) & 0.0 & 0.0 & -0.44(3) & 0.0 & 0.0 \tabularnewline
6.88(3) & 0.0 & 0.0 & -0.897(13) & 0.0 & 0.0 & -0.44(3) \tabularnewline
6.88(3) & -0.44(3) & 0.0 & 0.0 & 0.897(13) & 0.0 & 0.0 \tabularnewline
10.31(5) & 0 & 0 & 0 & 0 & -1.0 & 0 \tabularnewline
10.31(5) & 0 & -1.0 & 0 & 0 & 0 & 0 \tabularnewline
\end{tabular}\end{ruledtabular}
\label{tab:Eigenvectors1}
\end{table}

\begin{table}[h]
\caption{Eigenvectors and Eigenvalues of the fitted CeSiI Hamiltonian for fitted model 2 (Table \ref{tab:CEFparams}).}
\begin{ruledtabular}
\begin{tabular}{c|cccccc}
E (meV) &$| -\frac{5}{2}\rangle$ & $| -\frac{3}{2}\rangle$ & $| -\frac{1}{2}\rangle$ & $| \frac{1}{2}\rangle$ & $| \frac{3}{2}\rangle$ & $| \frac{5}{2}\rangle$ \tabularnewline
 \hline 
0.0 & 0.0 & 0.0 & -0.66(5) & 0.0 & 0.0 & 0.75(5) \tabularnewline
0.0 & -0.75(5) & 0.0 & 0.0 & -0.66(5) & 0.0 & 0.0 \tabularnewline
6.952(9) & 0.0 & 0.0 & 0.0 & 0.0 & -1.0 & 0.0 \tabularnewline
6.952(9) & 0.0 & -1.0 & 0.0 & 0.0 & 0.0 & 0.0 \tabularnewline
10.43(4) & 0.0 & 0.0 & -0.75(5) & 0.0 & 0.0 & -0.66(5) \tabularnewline
10.43(4) & -0.66(5) & 0.0 & 0.0 & 0.75(5) & 0.0 & 0.0 \tabularnewline
\end{tabular}\end{ruledtabular}
\label{tab:Eigenvectors2}
\end{table}

\begin{table}[h]
\caption{Eigenvectors and Eigenvalues of the fitted CeSiI Hamiltonian for fitted model 3 (Table \ref{tab:CEFparams}).}
\begin{ruledtabular}
\begin{tabular}{c|cccccc}
E (meV) &$| -\frac{5}{2}\rangle$ & $| -\frac{3}{2}\rangle$ & $| -\frac{1}{2}\rangle$ & $| \frac{1}{2}\rangle$ & $| \frac{3}{2}\rangle$ & $| \frac{5}{2}\rangle$ \tabularnewline
 \hline 
0.0 & 0 & -1.0 & 0 & 0 & 0 & 0 \tabularnewline
0.0 & 0 & 0 & 0 & 0 & -1.0 & 0 \tabularnewline
6.88(2) & 0.0 & 0.0 & 0.45(9) & 0.0 & 0.0 & -0.89(5) \tabularnewline
6.88(2) & -0.89(5) & 0.0 & 0.0 & -0.45(9) & 0.0 & 0.0 \tabularnewline
10.31(4) & 0.45(9) & 0.0 & 0.0 & -0.89(5) & 0.0 & 0.0 \tabularnewline
10.31(4) & 0.0 & 0.0 & -0.89(5) & 0.0 & 0.0 & -0.45(9) \tabularnewline
\end{tabular}\end{ruledtabular}
\label{tab:Eigenvectors3}
\end{table}

By comparing with the point charge model $B_n^m$ (especially $B_2^0$), one might be tempted to identify model 1 as the correct model, with a mixture of $J_z = \pm 5/2$ and $J_z = \pm 1/2$. However, the ground state determined from x-ray spectroscopy matches to within uncertainty only with model 2---which happens to be the furthest from the point charge model, and has the worst best fit $\chi^2_{red}$. 
This situation, where CEF neutron spectra alone give an underdetermined fit, is not uncommon \cite{PhysRevB.101.144432,Scheie_2022_CEFuncertainty}. Fortunately, in this case we are able to use a multi-probe approach to fully constrain the ground state and the Hamiltonian, resulting in the eigenspectrum in Table \ref{tab:Eigenvectors2}. 

And thus the ground state single ion doublet in CeSiI is
$\psi_0{\pm} = 0.66(5)|\mp \frac{1}{2} \rangle \pm 0.75(5)| \pm \frac{5}{2} \rangle $. Its $g$-tensor values are 
$g_{xx} = g_{yy} = 1.1 \pm 0.2$, 
$g_{zz} = 2.0 \pm 0.4$. 
This is a nearly isotropic effective spin which includes significant $J_{\pm}$ quantum tunneling terms within the doublet. This allows for the cycloidal order reported here as well as the strong quantum fluctuations and Goldstone magnon modes. 

We attempted a self-consistent fit with both the neutron and XAS data, whereby the $\alpha$ value derived from the XAS model is incorporated into a global $\chi^2$ function based on the computed $\alpha$ from the fitted CEF Hamiltonian. Because it is not straightforward to combine the uncertainties from these two types of measurements (given the more indirect nature of the XAS modeling), we assigned the $\chi^2$ roughly equal weights for the two different measurements. This yielded a fitted ground state wavefunction 
$\psi_0{\pm} = 0.67(2)|\mp \frac{1}{2} \rangle \pm 0.74(2)| \pm \frac{5}{2} \rangle $
and best fit CEF parameters $B_{2}^{0}=(0.04 \pm 0.03)$~meV, 
$B_{4}^{0}=(-0.0069 \pm 0.0004)$~meV, and $B_{4}^{3}= (0.544 \pm 0.006)$~meV. The derived $g$-tensor values are 
$g_{xx} = g_{yy} = 1.42^{+0.06}_{-0.08}$, 
$g_{zz} = 1.45^{+0.17}_{-0.11}$. 
Qualitatively, the picture is the same as the model 2 fit, though some quantities change. However, because the combination of the $\chi^2$ values is based on an arbitrary weight of the XAS data, we are unable to confidently claim the combined fitted ground state represents the best statistical knowledge. Thus for the reported values in the main text we merely use XAS to adjudicate between the fitted models from neutron data---which still identifies a unique ground state for the CeSiI single-ion wavefunction. 

Finally, we  note that at 60~K the CEF modes noticeably shift to lower energies, shown most clearly in Fig.~3(c) of the main text. This could be partly due to thermal expansion in the lattice moving the ligands further from the Ce ion, and partly thermal population into itinerant bands reducing the effective charges. Regardless, the relative intensities do not change much, and thus we expect the CEF eigenvectors at elevated temperatures to be similar to those in Table \ref{tab:Eigenvectors3}. 

\subsection*{Phonon broadening of CEF modes}

As is evident from main text Fig. 2, the crystal field modes are noticeably broader than the energy resolution of the neutron spectromemter, especially the 10.3~meV mode. CEF-mode broadening at low temperature can be caused by phonon coupling \cite{Scheie_CEFphonon_2025} and Kondo interactions \cite{Lopes_1988}, both of which are present in CeSiI. 
Disentangling these effects is not straightforward, but here we estimate the effects of CEF-phonon broadening using the method in Ref. \cite{Scheie_CEFphonon_2025}. 

\begin{figure*}
    \includegraphics[width=\textwidth]{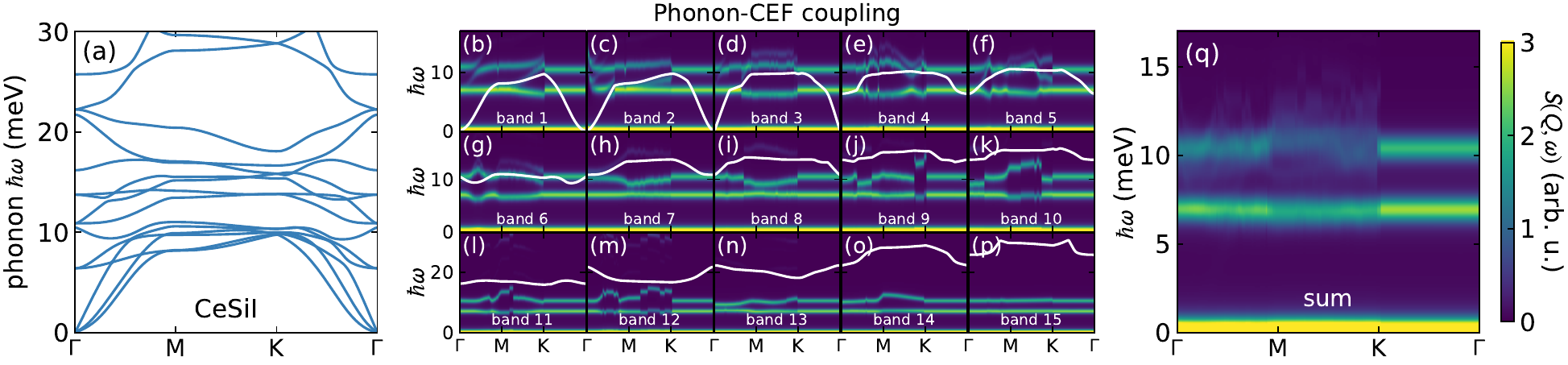} 
    \caption{Computed DFT phonon broadening for CeSiI. Panel (a) shows the DFT calculated phonon spectrum, which includes 18 phonon bands. Panels (b) through (p) show the calculated CEF spectrum for the lowest 15 phonons (the three highest energy phonons have negligible influence on the phonon modes). The colormap shows the momentum-dependent calculated CEF spectra, and the white line indicates the phonon dispersion along the $\Gamma = (0,0,0)$, $M=(1/2,0,0)$, $M=(1/3,1/3,0)$, $\Gamma$ path. Where the phonon  modes are close in energy, there is some numerical instability which causes some jaggedness in the phonon distortion versus momentum. However, the overall effect is clear in panel (q) which shows the sum of the spectra in panels (b) through (p), showing noticeable broadening in both modes, but especially in the higher-energy mode.}
    \label{fig:CSI_CEF_phonons}
\end{figure*}

We first computed the phonon spectrum using density functional theory (DFT) using the \emph{Vienna ab Initio Simulation Package} (VASP) version 5.4.4~\cite{kresse1993ab, kresse1996efficient, kresse1996efficiency, kresse1994norm} using the projector augmented wave (PAW)~\cite{blochl1994projector, kresse1999ultrasoft} method and the PBE exchange-correlation functional~\cite{perdew1996generalized}.
A $\Gamma$-centered Monkhorst-Pack~\cite{PhysRevB.13.5188} k-mesh was used for all calculations. Phonons were calculated using the Phonopy software~\cite{phonopy-phono3py-JPCM,phonopy-phono3py-JPSJ}. We used the DFT-D3 method of Grimme with zero-damping function for the van der Waal-dispersion energy-correction term to describe the van der Waal interactions in this layered material.~\cite{grimme2010consistent}  We used a PAW potential that treats the Ce 4$f^1$ electron as part of the core (valence configuration $5s^2 5p^6 6s^2 5d^1$). The PAW potentials used for all calculations (\texttt{Ce\_3}, \texttt{I}, and \texttt{Si}) have \texttt{ENMAX} values 176.506, 175.647, and 245.345 eV  respectively, so that 250 eV should be the minimum plane wave energy cutoff ($E_\mathrm{cut}$) used. We tested convergence of the total energy with respect to $E_\mathrm{cut}$ for the 6 atom nonmagnetic primitive unit cell (space group P$\bar{3}$m1 \#164, Wyckoff positions 2c (Ce) and 2d [Si and I]). We found convergence of 0.1 meV/atom at $E_\mathrm{cut} = 450$ eV. The experimental primitive lattice constants we use are {$a=b=4.170\;\mathrm{\AA}$ and $c=11.676\;\mathrm{\AA}$}. We found that a $4\times4\times1$ supercell of the primitive cell (96 atoms) and a k-mesh of $10\times10\times4$ were sufficient to obtain converged forces and phonon dispersions. The calculated phonon bands are shown in Fig. \ref{fig:CSI_CEF_phonons}(a). 

We then computed the CEF-phonon coupling using a point-charge model to estimate the phonon distortion $\mathcal{O}_{\mu}$ of the ligand environment on the CEF levels \cite{Scheie_CEFphonon_2025}. 
We write the total Hamiltonian as a linear combination of the Stevens and distortion operators 
\begin{equation}
    \mathcal{H} = \mathcal{H}_{CEF} + \sum_\mu \left[ \hbar \omega_\mu \left(a_{\mu}^{\dagger} a_{\mu} + \frac{1}{2} \right) + (a_{\mu} + a_{\mu}^{\dagger} ) \mathcal{O}_{\mu} \right] 
    \label{eq:CEF-phonon}
\end{equation}
\cite{Thalameier_1982,Thalmeier_1984} 
where $\mathcal{H}_{CEF}$ is the un-distorted CEF Hamiltonian (Eq. \ref{eq:CEF_H}), $a_{\mu}$ and $a_{\mu}^{\dagger}$ are the phonon creation and annihilation operators of phonon $\mu$ at energy $\hbar \omega_\mu$, and $\mathcal{O}_{\mu}$ is the phonon distortion operator.
Diagonalizing the Hamiltonian in Eq. \ref{eq:CEF-phonon} with the phonons one-by-one yields the spectra in Fig. \ref{fig:CSI_CEF_phonons}(b)-(p). Many of these phonons strongly couple to the two CEF levels, but the strongest coupling is to the 10.3~meV $J_m= \pm \frac{3}{2}$ level. This is most clearly seen in Fig. \ref{fig:CSI_CEF_phonons}(q), which shows the summed CEF spectra over the $\Gamma \rightarrow M \rightarrow K \rightarrow \Gamma$. This is not the same as a powder-average, which would require summing over a volume of reciprocal space vectors. Nevertheless, this clearly shows (i) non-negligible CEF-phonon interactions which cause broadening in the observed CEF levels, and (ii) a more dramatic broadening of the 10.3~meV $J_m= \pm \frac{3}{2}$ mode. 

The Kondo coupling may add to this broadening, but the energy-overlap of the CEF modes with many phonon modes strongly suggests that the observed CEF broadening is in large part (if not mostly) due to CEF-phonon interactions. The fact that this simple point-charge model gets the level of broadening correct for the higher and lower excited modes is also confirmation of the CEF eigenspectrum derived above. 
Because of the approximate nature of this calculation we do not incorporate the CEF-phonon coupling into the fitted CEF model and instead treat the broadening as a phenomenological parameter---and still arrive at consistent results between neutrons and X-ray scattering. 

\clearpage
\newpage

\section{Crystal electric fields studied with X-ray absorption spectroscopy}
To determine the ground state symmetry of CeSiI, we performed X-ray absorption spectroscopy (XAS) and resonant inelastic X-ray scattering (RIXS) measurements at the Ce $M_5$-edge ($3d \to 4f$). Measurements were performed at the 2-ID (SIX) beamline at the National Synchrotron Light Source II, Brookhaven National Laboratory~\cite{2016_Dvorak_RevSciInstrum}. XAS measurements were recorded in total electron yield (TEY). We used single crystal CeSiI samples cleaved in-situ in ultrahigh vacuum ($P<10^{-8}\,$mbar) at room temperature.

The XAS curves in both theory and experiment were normalized by the polarization-averaged integral of the Ce M$_5$ absorption. The x-ray linear dichroism (XLD) is extracted by the difference of the two independent polarizations after normalization. The theoretical curves were convolved with an energy-dependent intermediate-state (Lorentzian) lifetime broadening $\Gamma(\omega)$ between $\Gamma = 0.3$ eV in the pre-edge to $\Gamma = 0.9$ eV at the peak of the $M_5$ absorption to best fit to the polarization averaged XAS. For fitting the XLD, we use a thermally averaged summation of the contributions from the three Kramer’s doublets with weights chosen to match the integral of the absolute value of the experimental XLD. The ground state is determined to be of the form $|\psi\rangle = \alpha \cdot |\pm 5/2 \rangle + \beta \cdot |\pm 1/2 \rangle$  with $\alpha^2 = 0.61(3)$ and $\beta^2 = 0.39(3)$. The error bars are determined averaging over fits performed for four independent measurements on three separate samples.

Along with the energy levels determined by the INS spectrum, the XLD results are consistent with two sets of CEF parameters. The first has the pure $J_z = \pm 3/2$ as the first excited state with 
$B_2^0=-0.13$~meV, $B_4^0 = -0.0050$~meV and $|B_4^{\pm 3}| = 0.5321$~meV, while the second has the pure $J_z = \pm 3/2$ as the second excited state with $B_2^0=-0.155$~meV, $B_4^0 = -0.0237$~meV and $|B_4^{\pm 3}| = 0.3554$~meV (Stevens formalism). 
The first set of parameters is close to the second parameter set (model 2, ‘GS2’). We thus conclude that this set of parameters and level sequence are most consistent with the totality of the data.

\begin{figure}

    \includegraphics[width=\textwidth]{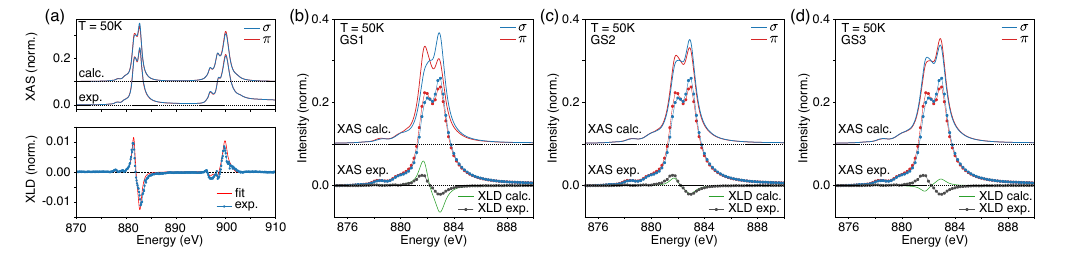} 
    \caption{(a) Ce $M_{4,5}$-edge XLD in TEY on a second sample at T = 50 K and at $\theta = 45^\circ$ with incident $\sigma$ (blue) and $\pi$ (red) polarizations and (b) the corresponding XLD spectrum. Data are compared to an optimized fit [upper curves in (a), red solid curve in (b)]. (b-d) Comparison of XAS and XLD spectra from different optimized fits to the inelastic neutron spectra (GS1, GS2 and GS3). Experimental data are the same as in main text Fig.~2 and shown as data points. Solid curves are the calculated XAS/XLD spectra for the different sets of crystal field parameters determined by fits to the INS data (see text for details). } 
    \label{fig:FigureXAS}
\end{figure}

\clearpage
\newpage

\section{Linear spin wave theory}
Linear spin wave theory (LSWT) calculations for Ce$^{3+}$ ($S=\frac{1}{2}$, $g=\frac{6}{7}$) were carried out with the Julia based package \texttt{Sunny}~\cite{2025_Dahlbom_}. 
For simulations on Hamiltonians including exclusively isotropic exchange, we worked with a system size that corresponds to the crystallographic unit cell. Here, \texttt{Sunny} simulations were based on the rotating frame formalism~\cite{2015_Toth_JPhysCondensMatter}. For Hamiltonians including anisotropic exchange or DM interactions, we used larger unit cells, as specified in each case.

\clearpage
\newpage

\section{Dzyaloshinskii-Moriya interactions in $\bf CeSiI$}

\begin{figure}

    \includegraphics[]{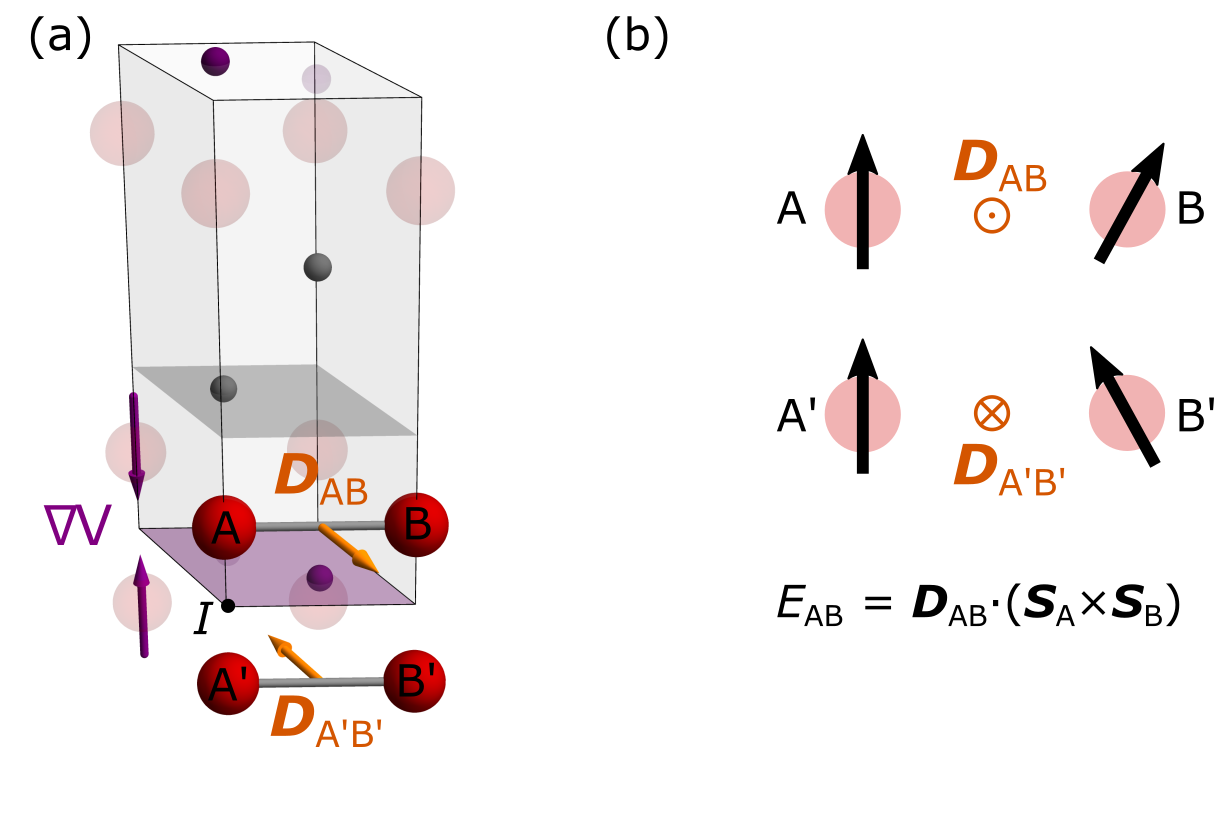} 
    \caption{Nature of DM interactions in CeSiI. (a) The local envoronment of Ce ions is noncentrosymmetric, thereby enabling DM interactions even though the global symmetry is centrosymmetric. For bonds perpendicuilar to the $c$-axis, such as $\left\langle100\right\rangle$, the stacking sequence of I and Si layers leads to a local electric field, $\nabla V$, that is inverted on the nearest-neihgbor Ce layer. Therefore, the resulting DM vectors between two bonds AB and A$'$B$'$, DM vectors are also pointing in opposite direction. (b) In this specific case, DM vector for AB implies that the DM-energy $E_{AB}$ is minimized for a cycloidal spin rotation between A and B. For the nearest- neighbor bonds due to the opposite sign of the DM vector, a cycloidal rotation of opposite rotational sense is favored.} 
    \label{fig:FigureDMVectors2}
\end{figure}

According to the rules for antisymmetric exchange established by Moriya, cf. Ref.~\cite{moriyaAnisotropicSuperexchangeInteraction1960a}, DM vectors may be allowed between magnetic ions, if the center of their connecting bond does not represent an inversion symmetry center. In CeSiI, this inversion symmetry is broken for any bond within a $(001)$ layer (i.e. perpendicular to $[001]$) due to the noncentrosymmetric stacking of Si and I layers around each Ce layer (see Fig.\ref{fig:FigureDMVectors2}(a)). As argued in the following, DM vectors in CeSiI favor opposite sense of spin rotation on neighboring Ce sublattices (a Ce sublattices is constituted by the Ce sites that are related by translation).

First of all, we consider a bond AB along $\left\langle100\right\rangle$ on one Ce sublattice. The inversion related bond B'A' on the nearest-neighboring Ce layer belongs to another Ce sublattice. The respective DM vectors are antiparallel, $\bm{D}_{AB}=-\bm{D}_{A'B'}$, and therefore they favor opposite rotational sense on neighboring Ce-layers.

This may be alternatively understood in the picture of Rashba spin--orbit interactions: Local electric fields created by Si and I layers are inverted on neighboring Ce-layers. A putative exchange particle with effective momentum $\bm{k}\parallel \bm{R}_{AB}=
a\cdot \hat{e}_x$ moving from site A to B experiences spin-orbit force $\sim \bm{k}\times\nabla V$~\cite{2005_Kane_PhysRevLettb} and therefore has opposite direction on nearest neighboring Ce layers (moving from A$'$ to B$'$). In the spin Hamiltonian this translates to antiparallel DM vectors on bonds AB and A$'$B$'$. 

We may assume more generally a bond AB on one Cerium sublattice. Inversion operator $I$ maps AB on a bond B'A' on the second Ce sublattice. As before, these two DM vectors are antiparallel and therefore favor opposite rotational sense.


As argued above, bonds perpendicular to $[001]$ may host DM interactions. For the shortest bonds with $c$ axis components that connect nearest-neighbor Ce layers, DM interactions are forbidden due to inversion symmetry.
For bonds connecting to a second nearest neighboring layer (with $\Delta a_3=c$), inversion symmetry is instead broken and DM vectors may be finite. The shortest bond that has a finite $c$-axis component and that further permits DM interactions is the $18$th shortest d bond of length $12.39\,$\AA (see also table~\ref{tab:NearestNeighborBonds}).

\clearpage
\newpage

\section{Steps in Magnetization}

\begin{figure}
    \includegraphics[]{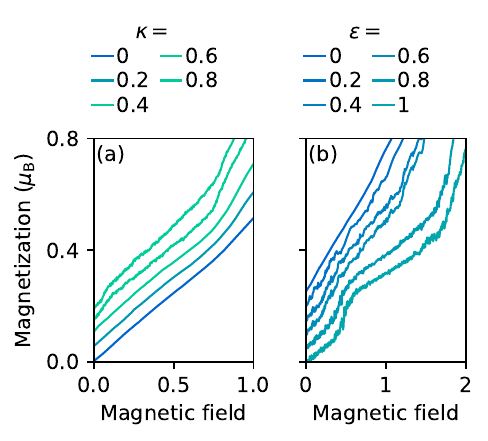} 
    \caption{Steps in magnetization as a function of field. The field dependence of magnetization, as calculated with Langevin Dynamics, displays almost linear field dependence for the isotropic Heisenberg model considered in the text. However, both (a) anisotropy and (b) DM interactions result in step-like magnetization. For (a), we considered single-ion anisotropy, $-\kappa\cdot {S^z}^2$ and for (b) DM vector $\epsilon\cdot(0.066,0.018,0.032)$ along the bond $[-1,0,-1]$, $\epsilon\cdot(0,-0.6,0.1)$ along $[-1,0,0]$, and $\epsilon\cdot(0.6,0,0)$ along $[1,2,0]$.}  
    \label{fig:FigureMagnetizationSI}
\end{figure}

The model presented in the main text (Eq.~2 with exchange couplings $J_1$, $J_2$, $J_4$, $J_5$--$J_{12}$) results in linear magnetization as a function of field without the steps seen experimentally~\cite{posey2024CeSiI}.
To illustrate this, we performed finite temperature simulations ($T=0.08\,$K) using Langevin dynamics, a variation of Landau-Lifshitz dynamics, implemented in \texttt{Sunny}~\cite{2022_Dahlbom_PhysRevB}.
The field dependence of magnetization is presented in Fig.~\ref{fig:FigureMagnetizationSI} (blue curve, $\kappa=0$).

As shown in the following, both single ion anisotropy and DM interactions can lead to a step-like magnetization. However, relatively strong magnitudes of these interactions are required to produce steps.
We first consider easy-axis single-ion anisotropy, represented by the term $-\kappa S_{z}^2$ in the Hamiltonian. Results for different magnitudes of $\kappa$ are presented in (a) and show clear steps.

In addition, we considered DM interactions as represented in the Hamiltonian by antisymmetric matrices:
\begin{align}
    A_{i\nu,j\mu} = \begin{pmatrix} 0 & +I & -H\\ -I & 0 & +G\\H&-G&0 \end{pmatrix} \, .
\end{align}
For bonds along $[100]$ the components translate directly to the DM vector $\bm{D}_{i\nu,j\mu}=(G,H,I)$. (b) presents the results for the following DM vectors: $\epsilon\cdot(0.066,0.018,0.032)$ along the bond $[-1,0,-1]$, $\epsilon\cdot(0,-0.6,0.1)$ along $[-1,0,0]$, and $\epsilon\cdot(0.6,0,0)$ along $[1,2,0]$. 

Steps are observed, however only for comparatively strong DM interactions reaching components of the order 0.5\,meV or for anisotropies of the order 0.8\,meV. Both cases are not supported by our neutron data.

\clearpage
\newpage

\section{Effective Hamiltonian for magnons in CeSiI}

In the main text, we presented an effective spin Hamiltonian that accounts for the major features in the neutron spectroscopy data, which are (i) the Goldstone mode at the magnetic zone center (seen also in higher Brillouin zones), (ii) the acoustic spectrum with enhanced intensity that appears like a flat mode at around 0.6 meV, and (iii) optical magnons that are most enhanced on what looks like a flat mode at around 1.6 meV.

To design this effective Hamiltonian, as a first step, 
we determined the interactions that lead to the modulation perpendicular to the $c$-axis, $k_x=0.28$. Luttinger-Tisza (LT) Analysis~\cite{1974_Litvin_Physica} shows that nearest- and next-nearest neighbor intralayer exchange ($J_2$ and $J_4$, respectively) within the plane $(001)$ already account for this modulation, if $J_4/J_2=1.6$ is satisfied.

We calculated the LT exchange according to Eq.~2 of the main text, which corresponds to a $2\times2$ matrix $\mathrm{J}(\bm{q})$ with Eigenvalues $E_1(\bm{q})\leq E_2(\bm{q})$. In cases, where $J_1$ is the only connecting exchange between the two cerium sublattices, labelled $\nu=1$ and $2$, minimization of Eq.~2 of the main text can be traced back to the Hamiltonian of a single cerium sublattice.
As all interactions are isotropic, the strong LT condition in the LT method is automatically satisfied with the weak condition~\cite{1960_Lyons_PhysRev}, implying that the magnetic ordering vector of the ground state corresponds to a global minimum of $E_1(\bm{q})$. Simple numerical calculations show that for $J_4/J_2=1.6$ (and no other exchange interactions), the global minimum satisfies $k_x=0.28$.

\begin{figure}
    \includegraphics[]{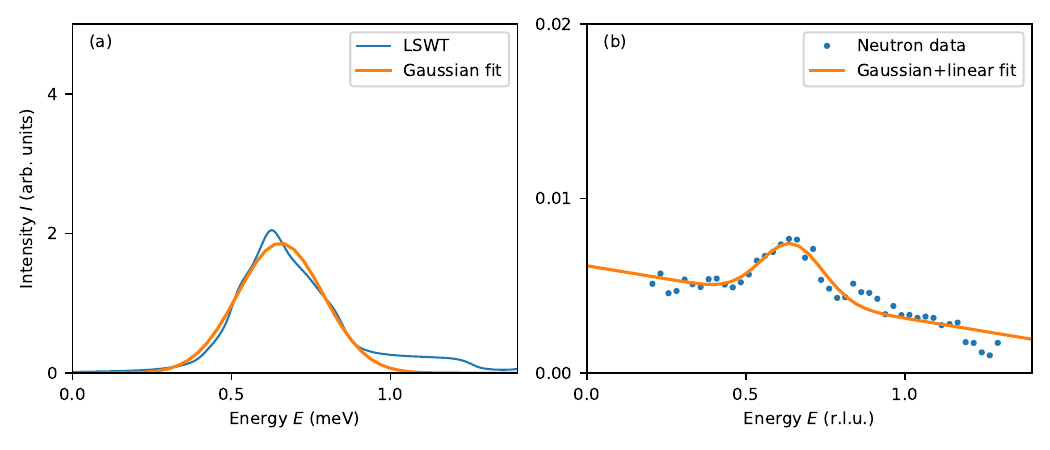} 
    \caption{Determination of the energy scale of $J_2$ and $J_4$. (a) Linecut through powder averaged $S(Q,\omega)$ at $Q=0.35\,\mathrm{\AA}^{-1}$, as calculated with linear spin wave theory (LSWT). The orange line shows a Gaussian fit. Here, $J_2=-0.39$ and $J_4=0.62$ were chosen sothat the maximum of the linecut appears at the same energy as in the experimental data [panel (b)]. (b) Linecut through neutron spectroscopy data at $Q=0.35\,\mathrm{\AA}^{-1}$. The orange line shows a fit with a Gaussian profile plus linear $Q$-dependence.
    } 
    \label{LinecutsJ2J4}
\end{figure}

To determine $J_4$ and $J_2$ on an absolute scale, we consider the linecut at $Q=0.35\,\mathrm{\AA}^{-1}$ through neutron spectroscopy data, which displays maximum intensity at $E=0.65\,$meV (see Fig.~\ref{LinecutsJ2J4}). 
The maximum in the simulated data is scaled by the absolute magnitude of $J_2$ and $J_4$. The maximum appears at $E=0.65\,$meV for $J_2=-0.39$ and $J_4=0.62$. (see Fig.~\ref{LinecutsJ2J4}). 

\begin{figure*}
    \includegraphics[width=1
    \textwidth]{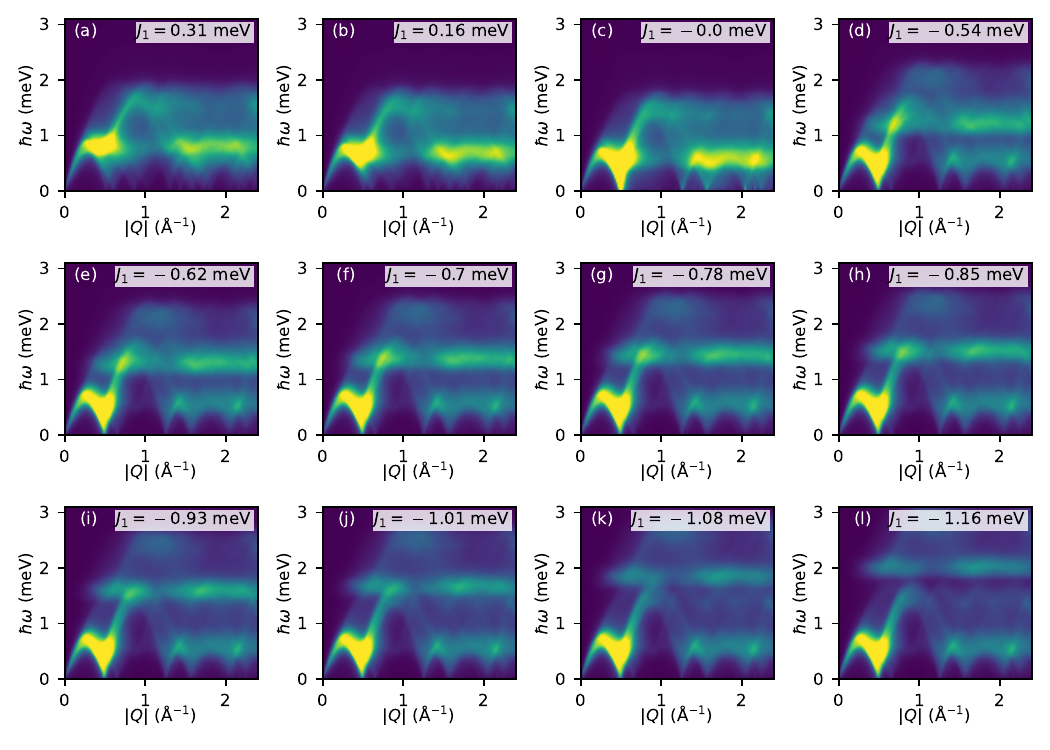} 
    \caption{Powder spectroscopy simulations for a Heisenberg Hamiltonian with $J_2=-0.39$, $J_4=0.62$, and different values of $J_1$. All other exchange terms were set to zero.
    } 
    \label{fig:CopmareJ1}
\end{figure*}

The spectrum of this minimal model, where only $J_2$ and $J_4$ are finite, is shown in Fig.~\ref{fig:CopmareJ1}(c). It reproduces the experimental spectrum of accoustic magnons, but not of the optical magnons. The flat horizontal signature due to optical magnons is pronounced for negative values of $J_1$ [Fig.~\ref{fig:CopmareJ1}(a-l)], which corresponds to ferromagnetic coupling, and matches the experimental data for $J_1=-0.93$ [panel (i)].

The Hamiltonian including exchange coupling $J_1$--$J_4$ represents a quasi two dimensional model that does not account for the spiral modulation along the $c$-axis, $k_z=0.19$. The ground state of this model corresponds to a cycloid. As $J_1$ is the dominant exchange in the system, a non-collinear canting of spins connected by a nearest neighbor bond is highly disfavored, implying that the cycloids on the two cerium sublattices need to have same rotational sense. As, in addition, $J_1$ is ferromagnetic, the two cycloids are rotating in phase, rather than phase shifted.

Here we further show that relatively weak interactions $J_5$--$J_{12}$ can result in the right three dimensional wave-vector, $\bm{k}$. Using global minimization and LT analysis, we found that $J_5=-0.20$, $J_{11}=-0.02$, and $J_{12}=0.08$, in addition to $J_1$--$J_4$, provide a global minimum of the LT exchange at $\bm{Q}=\bm{k}$.
In this case the Hamiltonian cannot be reduced to a single sublattice anymore, where LT analysis is known to be exact. However, we verified using \texttt{Sunny} that the ground state for this combination of interactions indeed possesses wave vector $\bm{k}$.

\begin{figure}
    \includegraphics[width=0.8\textwidth]{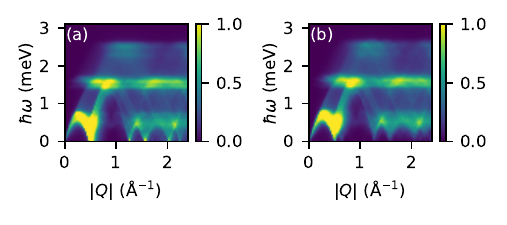} 
    \caption{Calculated powder spectrum for (a) a Heisenberg model with exchange couplings $J_1$-$J_4$ and (b) a Heisenberg model with exchange couplings $J_1$-$J_{12}$. Values for the $J$'s were chosen as specified in the text.
     } 
    \label{fig:Copmare2D3D}
\end{figure}

The exchange couplings of the model with $J_1$--$J_{12}$ are relatively weak and do not change the powder spectroscopy pattern, as demonstrated in Fig.~\ref{fig:Copmare2D3D}.
Panel (a) shows LSWT calculations for the quasi two dimensional model including exchange couplings $J_1$--$J_4$. (b) shows the model with $J_1$--$J_{12}$. Magnetic zone centers in the powder average appear at almost the same $Q$ as 2d and 3d wave vector differ only by ($|0.28\cdot b_1-\bm{Q}_0|=0.01\,\text{\AA}^{-1}$). This is due to the small length of the reciprocal $\mathbf{c}^*$ axis compared to the $\mathbf{a}^*$ axis.

%

%

As demonstrated with \texttt{Sunny}, a spiral state still provides a stable ground state of the full Hamiltonian in Eq.~2 (with exchange terms for the twelve shortest nearest-neighbor bonds $J_1$--$J_{12}$) with wave-vector $\bm{k}$. As $J_1$ is still the dominant inter sublattice exchange, cycloids on sublattices are still co-rotating for the full 3d model.

However, as the Hamiltonian only includes isotropic exchange, it is invariant under global spin rotation (SU(2) symmetry). Therefore it does not fix the plane of spiral rotation. Our model thus cannot distinguish between a cycloidal or helical spin rotation. 

However, neutron powder diffraction showed clearly that moments lie in the plane defined by $\mathbf{c}$ and \replaced{$\mathbf{b}^*$}{$\mathbf{a}^*$} (which is perpendicular to \replaced{$\mathbf{b}$}{$\mathbf{a}$}). The only ground state that is in agreement with neutron powder diffraction and Eq.~2 of the main text corresponds to the co-rotating cycloid shown in Fig.~1(c) of the main text. Therein, magnetic moments undergo a full rotation on a circle spanned by the vectors $\mathbf{c}$ and \replaced{$\mathbf{b}^*$}{$\mathbf{a}^*$}, with a periodicity of around 5 \added{lattice parameters $c$} and \replaced{3.1}{3.5} lattice parameters \added{$a$} along $[001]$ and \replaced{$\mathbf{b}^*$}{$a*$}, respectively. The texture in main text Fig. 1(c) represents one of three domains according to $120^{\circ}$ and $240^{\circ}$ rotations around the $c$-axis in trigonal symmetry. Modulations of the magnetic moment with larger amplitude along the easy $c$-axis, as indicated in neutron powder diffraction~\cite{2021_Okuma_PhysRevMater}, are not captured by our model, but do not affect our main results on the spin dynamics.

One final limitation on this model is that we are unable to find a ground state or compute the spectrum from a spin-density wave ground state. Thus, technically, we have not ruled this out as a possible magnetic ground state for CeSiI. However, the intense Goldstone mode still allows us to determine the dominant sign of the magnetic exchange for the nearest neighbor, and this will not be changed by a spin density wave ground state.

\newpage 
\clearpage

\section{Effect of anisotropic nearest neighbor exchange on powder spectroscopy}

Isotropic nearest neighbor exchange in Eq.~2 of the main text favor co-rotating spin spirals and leads to the best agreement between simulations and neutron data [see Fig.~3 panels (a) and (c) of the main text].

Counter-rotating spirals would instead be favored, if $J_1$ had a strongly Ising or XY-like character.
In the main text, we showed simulations with weak Ising or XY character with the exchange matrices:
\begin{align}
\mathrm{J}_{1,\mathrm{Ising}} = \begin{pmatrix}
0.5\cdot J_1 & 0 & 0\\
0 & 0.5\cdot J_1 & 0\\
0 & 0 & J_1 \end{pmatrix} \,\,\,\, \mathrm{and,} \,\mathrm{J}_{1,\mathrm{XY, Ising}} = \begin{pmatrix}
 J_1 & 0 & 0\\
0 &  J_1 & 0\\
0 & 0 & 0.5 \cdot J_1
\end{pmatrix}
\end{align}
These simulations were done in a supercell of 100 conventional unit cells along the $[100]$ direction.

As shown in the main text, the anisotropy with intermediate strength of nearest neighbor exchange already leads a gapped dispersion and a rather diffuse scattering with weak intensity below 0.5\,meV. This (and stronger anisotropies as required for a counter-rotating spiral) is/are not supported by our data.

\newpage 
\clearpage

\section{Effect of DMI on powder spectroscopy}

\begin{figure}[b!]
    \includegraphics[]{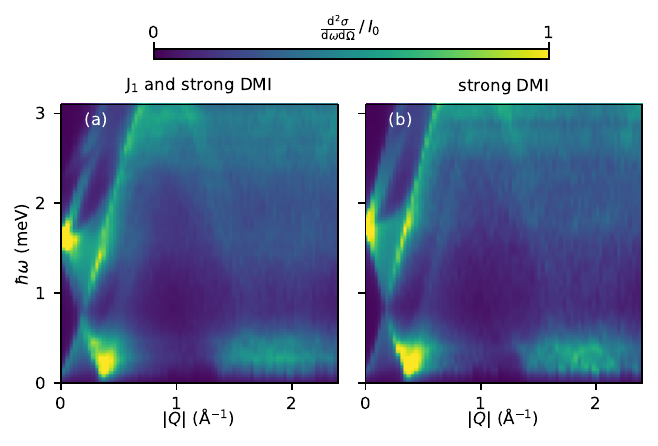} 
    \caption{Calculated powder spectra with DMI. (a) Shows calculations done for a model with parameters $J_1$--$J_4$ chosen as in the main text and with DM interactions characterized by the vector $(0,D,0)$ along the bond $[100]$ with $D=1.1926\,$meV. (b) Shows calculations done for a model with parameters $J_2$--$J_4$, with $J_1=0$, and with DM interactions characterized by the vector $(0,D,0)$ along the bond $[100]$ with $D=1.1926\,$meV. Calculations in \texttt{Sunny} were done in a system with cell size 10x1x1 as compared to the crystallographic unit cell.
    } 
    \label{fig:DMIExchange}
\end{figure}

Fig.~\ref{fig:DMIExchange} shows the powder spectroscopy pattern, calculated for two different Hamiltonians in that DM interactions stabilize the incommensurate modulation $k_x=0.28$.
(a) and (b) show calculations with $J_1=-0.93$ and $J_1=0$, respectively. In both cases, enhanced intensity at the GM is absent. The calculation in (a) corresponds to the calculation in Fig.~3(f) of the main text. We may therefore rule out the possibility that the incommensurate modulation $k_x$ is stabilized by DM interactions.

\newpage 
\clearpage

\newpage 
\clearpage

\section{Overview and labeling of bonds in the spin Hamiltonian}


Table~\ref{tab:NearestNeighborBonds} gives an overview of the 20 shortest nearest neighbor bonds between Ce sites in the crystal structure of CeSiI. In the main text, we used the notation $J_n$ for the exchange coupling along the $n$-th shortest bond. Alternatively, this coupling may be written as $J_{\bm{r}_i\nu,\bm{r}_i+\bm{t}_{ij}\mu}$ for the coupling connecting the Ce site in the unit cell $\bm{r}_i$ at the position $\nu$ with the Ce site in the unit cell $\bm{r}_j$ at the position $\mu$. $\nu$ and $\mu$ can have values $1$ and $2$, which denotes the two different Ce positions in the conventional hexagonal unit cell. $\bm{t}_{ij}=\bm{r}_j-\bm{r}_i$ denotes the translation vector between starting and targeting conventional unit cells.

\begin{table}[h]
\caption{Nearest neighbor bonds between Ce sites $i$ and $j$ in CeSiI sorted by length in increasing order. The first column labels the bonds with a integer. The third and fourth column provide the translation vector $\bm{t}_{ij}=\bm{r}_j-\bm{r}_i$ between the conventional unit cells containing $\bm{r}_i$ and $\bm{r}_j$, respectively. The fourth column provides the values $\nu$ and $\mu$, which indicate which of the two Ce positions in the conventional unit cell is chosen as starting and targeting point of the bond with index $n$ (see text for further details). The last column indicates, whether DM vectors are allowed by symmetry (+) or absent (.).}
\begin{ruledtabular}
\begin{tabular}{cc|cc|c}
Index $n$ & Length $L$& $\bm{t}_{ij}$ & $(\nu,\mu)$  & DM alw.
    \tabularnewline
 \hline 
1&3.97 & [0,0,-1] & (1,2)& .\tabularnewline
2&4.17 & [1,0,0] & (1,1)& +\tabularnewline
3&5.76 & [1,0,-1] & (1,2)& . \tabularnewline
4&7.23 & [1,2,0] & (1,1)& +\tabularnewline
5&7.70 & [0,0,0] & (1,2)& . \tabularnewline
6&8.24 & [1,2,-1] & (1,2)&. \tabularnewline

7&8.24 & [1,2,1] & (2,1)& .  \tabularnewline
8&8.34 & [2,0,0] & (1,1)& + \tabularnewline
9&8.76 & [1,0,0] & (1,2)& . \tabularnewline
10&9.24 & [2,0,-1] & (1,2)& . \tabularnewline
11&10.56 & [1,2,0] & (1,2)& . \tabularnewline
12&10.56 & [1,2,0] & (2,1)& .         \tabularnewline

13&11.03 & [2,-1,0] & (1,1)& + \tabularnewline
14&11.35 & [2,0,0] & (1,2)& . \tabularnewline
15&11.67 & [0,0,1] & (1,1)& . \tabularnewline
16&11.72 & [2,-1,-1] & (1,2)& . \tabularnewline
17&11.72 & [-2,1,-1] & (1,2)&  . \tabularnewline
18&12.39 & [1,0,1] & (1,1)&  +\tabularnewline
19&12.51 & [3,0,0] & (1,1)& +\tabularnewline
20&13.12 & [3,0,-1] & (1,2)& .\tabularnewline
\end{tabular}\end{ruledtabular}
\label{tab:NearestNeighborBonds}
\end{table}

\newpage 
\clearpage

\section{Quasielastic scattering}

Above the magnetic ordering temperature, we observed quasielastic scattering at energy  transfers below 0.5\,meV. In general, the quasielastic scattering may be momentum-dependent.
In Kondo lattice materials, the quasielastic scattering may be approximated in the context of local relaxation dynamics~\cite{1988_Grier_JPhysCSolidStatePhys,1981_Grier_PhysRevB,1981_Horn_PhysRevB}, which takes the $\bm{Q}$-independent form:
\begin{align}
    S(E,T) =  \frac{1}{1-\exp(-\beta E)}\cdot \frac{E}{E^2+\Gamma^2} \, ,
\end{align}
where $\Gamma$ is the quasielastic linewidth.

\begin{figure}[b!]
    \includegraphics[]{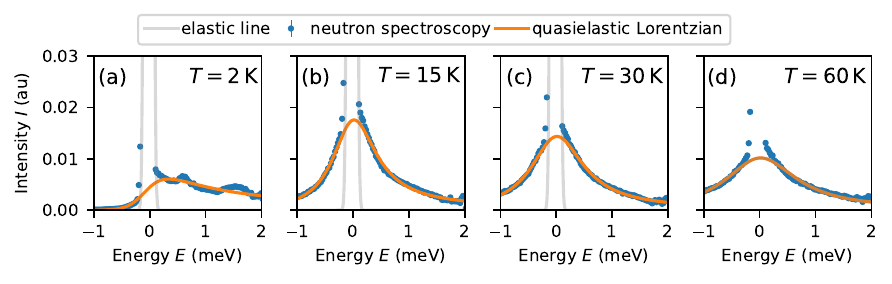} 
    \caption{Fit of the quasielastic scattering at different temperatures (a)-(d). Blue circles denote inelastic powder neutron spectroscopy data integrated over the momentum-range $0.4\,$\AA$^{-1}$\,$< Q<0.8\,$\AA$^{-1}$. The gray line denotes the elastic line. The orange line corresponds to the quasielastic profile of a single-particle relaxation process. 
    } 
    \label{fig:QuasielsticFits}
\end{figure}

\begin{figure}[b!]
    \includegraphics[]{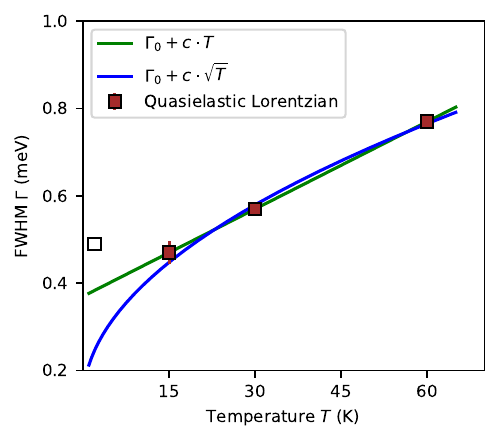} 
    \caption{Thermal variation of the quasielastic scattering linewidth. Squares denote the linewidth $\Gamma$, inferred through the fits in Fig.~\ref{fig:QuasielsticFits}. Green and blue lines, respectively, denote a linear and square-root temperature dependence fitted to the $\Gamma$ values in the paramagnetic state (squares that are filled red). The data point within the ordered state (filled white) was not considered for these fits.
    } 
    \label{fig:NeutronKondoTemperature}
\end{figure}

Accordingly, we fit the powder-averaged quasielastic scattering, integrated over momentum transfers $0.4\,$\AA$^{-1}$\,$< Q<0.8\,$\AA$^{-1}$ at different temperatures, as presented in Fig.~\ref{fig:QuasielsticFits}.
The thermal variation of the quasielastic scattering linewidth is presented in Fig.~\ref{fig:NeutronKondoTemperature}.
$\Gamma$ is of the order 0.4\,meV, which translates to a temperature of 4.6\,K.
The value is distinctively smaller than the characteristic temperature of Kondo coherence seen in resistivity ($T^{*}=50$\,K) and the Kondo temperature inferred from the Fano linewidth ($T_{\mathrm{K}}=74\,$K)~\cite{posey2024CeSiI}. 
Of course, these all being phenomenological models, there is no guarantee of internal consistency between the different measures of the Kondo energy scale. We thus look forward to the development of an effective microscopic model which can describe the energy scales of CeSiI in a self-consistent manner. 

\clearpage
\newpage